\documentclass[]{jfm} 
\usepackage{graphicx}
\usepackage{newtxtext}
\usepackage{newtxmath}
\usepackage{natbib}
\usepackage{hyperref}
\hypersetup{
    colorlinks = true,
    urlcolor   = blue,
    citecolor  = blue,
    allcolors  = blue
}

\newcommand{\RomanNumeralCaps}[1]
\linenumbers

\usepackage{xcolor}

\newcommand{\der}[2]{{d #1 }/{ d #2 } }
\newcommand{\Der}[2]{\frac {d #1 } { d #2 } }

\newcommand{\derp}[2]{\partial #1 / \partial #2  }

\newcommand{\Derp}[2]{\frac { \partial #1 } { \partial #2 } }

\newcommand{\dTtan}{\delta T_{\mu,\rho}}

\newcommand{\Anu}{A_\nu}
\newcommand{\dTnu}{\delta T_\nu}
\newcommand{\dTmurho}{\delta T_{\mu,\rho}}
\newcommand{\epsT}{\varepsilon_T}

\newcommand{\dPhi}{\Delta \Phi}
\newcommand{\dPhiz}{\Delta \Phi_0}

\newcommand{\COt}{\mathrm{CO}_2}

\newcommand{\ubar}{\bar{u}}
\newcommand{\mubar}{\bar{\mu}}
\newcommand{\rhobar}{\bar{\rho}}
\newcommand{\Tbar}{\bar{T}}

\newcommand{\du}{\Delta u}
\newcommand{\drho}{\Delta \rho}

\definecolor{darkGreen}{RGB}{4, 87, 1}
\definecolor{persoRed}{RGB}{232, 9, 9}

\shorttitle{Instability in strongly stratified plane Couette flow} 
\shortauthor{B. Bugeat et al.}

\title{Instability in strongly stratified plane Couette flow with application to supercritical fluids}

\author{B. Bugeat\aff{1} 
  \corresp{\email{b.bugeat@tudelft.nl}},
 P. C. Boldini\aff{1},
 A. M. Hasan\aff{1}
\and R. Pecnik\aff{1}
}

\affiliation{\aff{1}Process and Energy Department, Delft University of Technology, Leeghwaterstraat 39,
2628 CB Delft, The Netherlands
}

\usepackage[normalem]{ulem}

\usepackage{xcolor}

\begin{document}
\maketitle

\begin{abstract}
This paper addresses the stability of plane Couette flow in the presence of strong density and viscosity stratifications. It demonstrates the existence of a generalised inflection point that satisfies the generalised Fj\o rtoft's criterion of instability when a minimum of kinematic viscosity is present in the base flow. The characteristic scales associated with this minimum are identified as the primary controlling parameters of the associated instability, regardless of the type of stratification. To support this finding, analytical stability models are derived in the long wave approximation using piecewise linear base flows. Numerical stability calculations are carried out to validate these models and to provide further information on the production of disturbance vorticity. All instabilities are interpreted as arising from the interaction between two vorticity waves.
Depending on the type of stratification, these two waves are produced by different physical mechanisms. When both strong density and viscosity stratifications are present, we show that they result from the concurrent action of shear and inertial baroclinic effects. The stability models developed for simple fluid models ultimately shed light on a recently observed unstable mode in supercritical fluids (Ren \etal, \textit{J. Fluid Mech.}, vol. 871, 2019, pp. 831-864), providing a quantitative prediction of the stability diagram and identifying the dominant mechanisms at play. Furthermore, our study suggests that the minimum of kinematic viscosity reached at the Widom line in these fluids is the leading cause of their instability. The existence of similar instabilities in different fluids and flows (e.g., miscible fluids) is finally discussed.
\end{abstract}


\section{Introduction}

\subsection{Strongly stratified flows}

The stability of stratified parallel shear flows, in which fluid properties such as density and viscosity vary in the direction perpendicular to that of the base flow, is a problem encountered in several geophysical phenomena (e.g., dynamics of the atmosphere and the ocean) and industrial applications.
For example, fluids operating at pressure and temperature in the region of the critical point, which are employed in chemical and mechanical engineering \citep{brunner2010applications,liu2019supercritical}, 
may exhibit large variations of viscosity and density in flows involving heat transfers.
Stratified flows can be examined in different regimes.
In this paper, the variations of fluid properties will not be restricted to small amplitudes, justifying the term of strongly stratified flow. 
Besides, gravity, and therefore buoyancy effects, will be neglected by assuming large Froude numbers; more details on the flow assumptions will be given in section \ref{sec.intro_obj}. 
Density variations can nevertheless play a significant role in the flow dynamics through inertial effects.
\cite{menkes1959stability} was perhaps the first to tackle the stability of such a flow, considering a hyperbolic tangent velocity profile with an exponential density stratification, which was found stabilising in this particular configuration.
\cite{soteriou1995effects} more comprehensively studied an incompressible mixing layer of two fluids of different densities.
Depending on the density ratio, the primary instability was shown to exhibit either weaker or larger growth rates, to have its phase speed shifted, and its non-linear development altered.
This last point was subsequently examined via secondary stability analysis \citep{reinaud2000baroclinic,fontane2008stability} and direct numerical simulation (DNS) \citep{almagro2017numerical}.
The mechanism responsible for the modified dynamics of this flow is the inertial baroclinic torque, which generates vorticity from misalignments between pressure and density gradients \citep{soteriou1995effects,reinaud1999numerical,dixit2010vortex}.
It is also at play in compressible flows \citep{lesshafft2007linear} but is classically neglected in buoyant flows modelled via the Boussinesq approximation, which ignores density variations in inertial terms \citep{drazin1958stability,guha2018inertial}.

Strong viscosity stratifications will also be central in our problem, greatly affecting the base flow profile.
Considering a parallel shear flow of two fluids of different viscosities separated by an interface, \cite{yih1967instability} showed that a long-wave instability exists at low Reynolds numbers. 
This instability does not require density gradients or surface tension effects: the jump in viscosity at the interface is sufficient to destabilise the flow.
\cite{hooper1983shear}, in a similar configuration, revealed that a short-wave instability also grows at low Reynolds numbers.
The mechanisms of these instabilities were discussed by \cite{hinch1984note} and \cite{charru2000phase}.
The effect of an interface of finite thickness was studied by \cite{ern2003stability}.
The authors recovered the presence of low-Reynolds instabilities and furthermore showed that certain thicknesses could induce larger growth rates than an infinitely small one.
Finally, another viscous instability exists at larger but finite Reynolds numbers \citep{hooper1987shear}.
It is fundamentally different from the previous one as its mechanism is not directly associated with the presence of the viscosity interface but, rather, of the wall.
A comprehensive review of these instabilities for different flow configurations can be found in \cite{govindarajan2014instabilities}.

Plane Couette flow, which is linearly modally stable in the absence of stratification, was studied by \cite{joseph1964variable} in the presence of viscous heating, inducing temperature gradients and hence viscosity stratification.
A linear inviscid instability was shown to develop if a liquid, rather than a gas, was considered.
This observation was linked to the viscosity law, which decreases with temperature in liquids but increases in gas.
While this result, as the authors themselves stressed, did not proceed from a rigorous stability analysis as the linearised energy equation was decoupled from hydrodynamic effects, this instability was recovered by numerical calculations in subsequent works \citep{sukanek1973stability,yueh1996linear}.
However, these studies did not consider density variations, which may arise when considering viscous heating in gases. 
\cite{duck1994linear} carried out a stability analysis of plane Couette in a fully compressible framework.
The authors mostly focused on acoustic instabilities appearing at supersonic Mach numbers, as also later studied by \cite{malik2008linear} and \cite{saikia2017effects}.
In addition to the acoustic modes, \cite{hu1998linear} recovered the existence of a viscous mode similar to that found in the aforementioned incompressible, viscosity-stratified studies.

\subsection{Recent developments in the hydrodynamics of supercritical fluids}

Research on the hydrodynamics of fluids exhibiting non-ideal thermodynamic behaviour is actively progressing. 
A great deal of attention has recently been directed to understanding how the properties of these fluids affect turbulence, in particular turbulent heat transfer \citep{yoo2013turbulent}.
Recent studies have investigated the statistics of turbulence in different shear flows by means of DNS, for example in channel \citep{nemati2015mean,patel2016JFM,sciacovelli2017direct}, pipe \citep{peeters2016turbulence,he2021turbulence}, jet \citep{sharan2021investigation} or flat-plate boundary layer flows \citep{kawai2019heated,sciacovelli2020numerical}.
However, little is known about stability and transition to turbulence in these fluids \citep{robinet2019instabilities}.

\cite{gloerfelt2020dense} examined the linear stability of dense gas at large Mach numbers.
Due to the large heat capacity of these fluids, very weak temperature gradients were observed and nearly incompressible velocity profiles were recovered.
The authors showed the stabilisation of the viscous mode and the existence of radiating supersonic instabilities.
From a different perspective, \cite{ren2019linear} carried out a linear stability analysis of supercritical fluids in plane Poiseuille flow.
Having a lower heat capacity, significant viscous heating was present at reduced but non-negligible Mach numbers, generating temperature gradients in the base flow profile.
The authors concluded that non-ideal effects may induce larger destabilisation of the flow in terms of growth rate magnitude and critical Reynolds number.
In a subsequent study, \cite{ren2019boundary} explored the linear stability of supercritical $\COt$ in a flat-plate boundary layer flow.
As viscous heating was increased, a second unstable mode, in addition to the classical Tollmien-Schlichting (TS) wave, was observed.
This mode exhibits growth rates of more than one order of magnitude larger than the TS waves, which could imply new rapid modal routes of transition to turbulence in these fluids.
The authors rigorously showed that this mode was not linked to the Mack's modes \citep{mack1984boundary} found in high-speed boundary layer.
\cite{bugeat2022new} confirmed the inviscid nature of this instability and ruled out an acoustic origin.
Recently, \cite{ly2022} studied a binary compressible mixing layer at supercritical pressures and also found evidence of this instability, pointing out that its strength decreases as the reduced pressure is increased away from the critical point.
But much remains to be understood about this instability as the driving parameters and the physical mechanism remain unclear.

Importantly, \cite{ren2019boundary} observed that the additional mode only appears when the temperature profile of the base flow crosses the Widom line.
The concept of Widom line is specific to supercritical fluids.
It distinguishes the liquid-like from the gas-like region within the supercritical fluid domain. 
In each of these regions, fluid properties exhibit different behaviours \citep{simeoni2010widom}.
As such, the Widom line can be seen as the continuation of the coexistence line which separates the gas and liquid phases at sub-critical pressure, with the crucial difference that thermodynamic quantities smoothly vary across it \citep{banuti2015crossing,banuti2017similarity}.
These smooth variations can nonetheless exhibit remarkable behaviours.
At constant pressure, density and dynamic viscosity, as functions of temperature, feature strong gradients near the Widom line, while the kinematic viscosity can reach a minimum; see the introduction of \cite{ren2019linear} for more details on these behaviours and their implication for hydrodynamics.
Therefore, for a supercritical fluid operating at pressure and temperature near the Widom line, the presence of a temperature gradient in the flow leads to large density and viscosity variations; the flow is strongly stratified.

\subsection{Objectives, method and assumptions} \label{sec.intro_obj}

We aim to show that inviscid instabilities can be caused by the presence of a minimum of kinematic viscosity in strongly stratified shear flows, and that the scales associated with this minimum control the different properties of these instabilities. 
In particular, our objective is to provide evidence that the recently found unstable mode in supercritical fluids is related to the minimum of kinematic viscosity reached at the Widom line.
We also aim to identify the driving physical mechanisms at play in these instabilities.

A differentially heated plane Couette flow will be considered.
Three fluid models will first be examined, with different density and dynamic viscosity laws that strongly vary with temperature.
Different types of stratification will then be observed in the flow; however, the three fluid models are designed to all feature a minimum of kinematic viscosity.
In doing so, we aim to demonstrate the central role played by this minimum in the stability of the systems, regardless of the other property variations in the flow.
Using piecewise linear base flow approximations, analytical results will be derived by solving the Rayleigh equation in the presence of strong density gradients, which governs the inviscid linear stability of these flows.
A more realistic fluid model based on the van der Waals equation of state and diffusion laws at supercritical pressures will be used to ultimately discuss the instability in supercritical fluids.

The different hypotheses on the flow regime that we will consider in this work are summarised here.
No assumption regarding the magnitude of viscosity and density variations will be made.
Buoyancy will be ignored, but density variations will be retained in the inertial terms.
Acoustics will not be taken into account in order to remove potential ambiguities in the physical interpretation of the results with aforementioned acoustic instabilities.
The low-Mach approximation \citep{rehm1978equations,paolucci1982filtering} will be used.
As a result, no viscous heating will be at play; temperature gradients will be generated in the flow by boundary conditions.
Finally, only inviscid perturbations are considered.
Note that this is not inconsistent with the presence of viscosity-stratification effects in the base flow which, because it is parallel, is not affected by inertia.
The aforementioned instabilities induced by viscosity-stratification at low Reynolds number will therefore not be embedded in our analysis.
However, it should be kept in mind that a competition may take place at finite Reynolds numbers, where the inviscid instability is damped by viscous effects.

The paper is organised as follows.
The fluid and flow models, along with numerical procedures, are detailed in section \ref{sec.framework}.
The condition of existence of an inviscid instability in stratified plane Couette flow is examined in section \ref{sec.CouetteGIP}, leading to a criterion based on a minimum of kinematic viscosity.
The base flows of the fluid models are presented in section \ref{sec.baseflowsModel}.
Analytical stability results, based on piecewise linear models of these base flows, are derived in section \ref{sec.analyticalmodel}.
Comparison with numerical calculations is provided in section \ref{sec.LSA_calculations_models}.
The generation of disturbance vorticity by different physical mechanisms is also examined, and an interpretation of the different instabilities is proposed.
Section \ref{sec.resultsVdW} eventually focuses on the stability of a supercritical fluid.
A summary and a discussion on the application of these results to other fluids and flows conclude this paper (section \ref{sec.discussion}).

\section{Theoretical and numerical framework} \label{sec.framework}

    \subsection{Fluid models}\label{sec.fluid_models}

\begin{figure} 
            \centering
            \vspace{0.1cm}
            \includegraphics[angle = -0,trim=0 0 0 0,clip,width=0.97\linewidth]{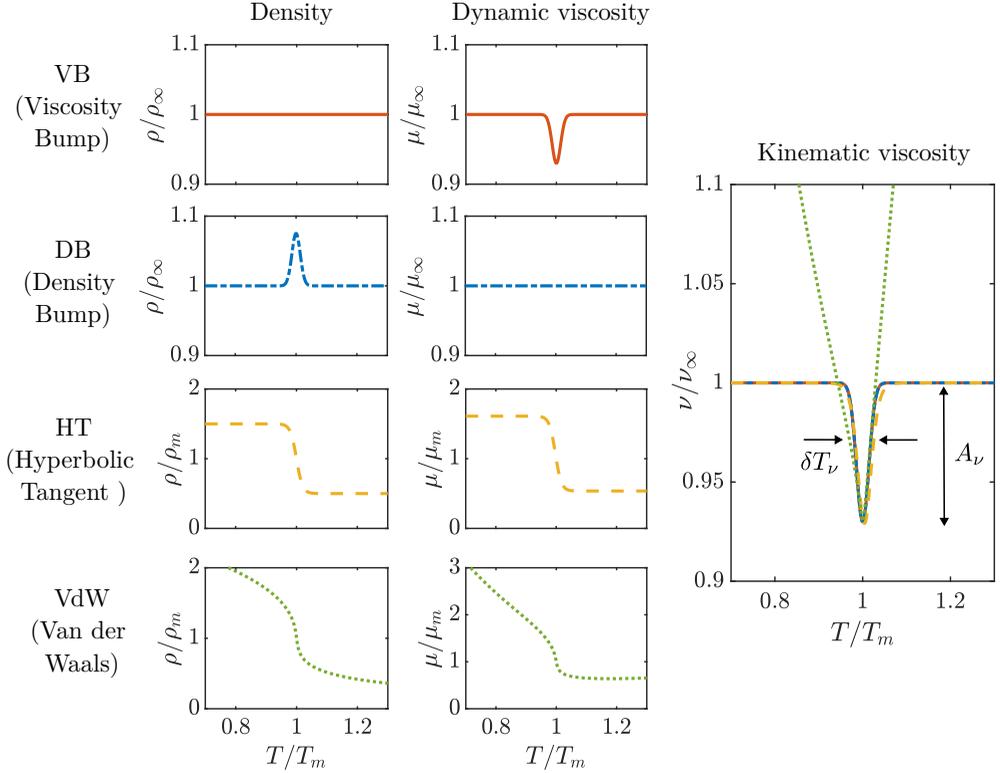}
            \captionsetup{justification=justified}        
            \caption{The four fluid models considered in this paper.
            The definition of $\nu_\infty$, used to normalise $\nu$, is given in appendix \ref{app.definition_VdW_dnu_Anu} in the particular case of fluid VdW.}
            \label{fig.Fluids}
        \end{figure}

        Four fluids will be considered throughout this paper, each of them being associated with a different equation of state and viscosity law. 
        However, they all share the common property of assuming an extremum of kinematic viscosity $\nu$ at a given temperature.
        Recalling that $\nu = \mu / \rho$, where $\mu$ and $\rho$ are the dynamic viscosity and the density, respectively, different ways to generate a minimum of $\nu$ can be imagined.
        Three theoretical fluid models will be used to control and study a restricted number of parameters.
        A fourth more realistic model for supercritical fluids, based on the van der Waals equation of state, will also be considered.
        A summary of the different fluids is provided in figure \ref{fig.Fluids} while a detailed description is given in the next subsections.
        
        \subsubsection{Fluid VB: Bump of dynamic viscosity with constant density} \label{sec.model1}

        In this model, density is assumed constant, while the viscosity is chosen to locally exhibit a bump at a temperature $T_m^*$, using a Gaussian function (in this paper, all dimensional quantities are noted using the superscript '*'). 
        Note that, in this case, the non-dimensional kinematic viscosity is equal to the non-dimensional dynamic viscosity, both reading
        \begin{equation} \label{eq.model1}
            \mu = \nu = 1 + A_{\nu} e^{ - \left( \frac{T - 1 }{ \delta T_\nu } \right)^2 }
        \end{equation}
        using $T_m^*$ as reference temperature.
        The reference viscosities are $\mu_\infty^*$ and $\nu_\infty^*$, which are the asymptotic values away from the bump.
        The parameter $A_\nu$ controls the amplitude of the bump, and its sign determines whether the kinematic viscosity admits a minimum ($A_\nu < 0$) or a maximum ($A_\nu > 0$). 
        The characteristic width of the bump is set through $\delta T_{\nu}$, which is again made non-dimensional using $T_m^*$.
        Finally, the thermal conductivity $\lambda$ is assumed to be constant.
        
        \subsubsection{Fluid DB: Bump of density with constant dynamic viscosity} \label{sec.model2}

        Inversely to fluid VB, $\mu$ is kept constant in fluid DB while a bump is introduced in the density profile.
        This bump is chosen such that the resulting kinematic viscosity has the same expression as in \eqref{eq.model1}.
        Hence, the density law simply reads
        \begin{equation} \label{eq.model2}
            \rho = \frac{1}{ 1 + A_{\nu} e^{ - \left( \frac{T - 1 }{ \delta T_\nu } \right)^2 } }
        \end{equation}

        \noindent and since, in this case, $\nu = 1/ \rho$, $\nu$ is the same as in fluid VB. 
        The conductivity $\lambda$ is again chosen to be constant.

        \subsubsection{Fluid HT: Hyperbolic tangent laws} \label{sec.model3}

        In fluid HT, thermal conductivity is also kept constant, while dynamic viscosity and density are now both allowed to vary according to hyperbolic tangent laws.
        In order to generate an extremum of kinematic viscosity, a small shift is introduced between the two hyperbolic tangents, controlled by the non-dimensional parameter $\varepsilon_T$.
        This choice is inspired by supercritical fluids and represents an attempt to mimic some of their features in the vicinity of the pseudo-boiling region.
        This will be discussed in more detail in section \ref{sec.modelVdW} after the supercritical fluid laws are introduced.
        The non-dimensional governing laws for fluid HT are formally written as
        \begin{align}  
            \rho &= 1 - \gamma \tanh \left[ \frac{ T - 1 }{ \delta T_{\mu,\rho} } \right] , \label{eq.model3rho} \\            
            \mu  &= 1 - \gamma \tanh \left[ \frac{ T - (1+\varepsilon_T) }{ \delta T_{\mu,\rho} } \right] . \label{eq.model3mu}
        \end{align}

        \noindent The reference temperature $T_m^*$ is here defined as the point of anti-symmetry of the density profile. 
        The density at $T=1$ and the viscosity at $T=1+\varepsilon_T$ are used as the reference scales.
        The parameter $\gamma$ controls the jump of density and dynamic viscosity while $\dTtan$ sets the temperature range over which this jump takes place.
        The fluid properties are shown in figure~\ref{fig.Fluids}, where it is verified that the kinematic viscosity admits a minimum around $T=1$.
        By analogy with fluid VB and DB, it is possible to estimate the amplitude $A_\nu$ of this minimum, as well as the characteristic width $\delta T_\nu$ of the temperature range onto which it occurs.
        The following relations will be used in this paper:
        \begin{align}
            A_\nu &=\frac{\gamma \varepsilon_T}{\delta T_{\mu,\rho}} , \label{eq.scalingAnu_tanh} \\
            \dTnu &= 1.2 \> \delta T_{\mu,\rho} .
        \end{align}
        The derivation and verification of these expressions are detailed in appendix \ref{app.FluidHT}.

        \subsubsection{Fluid VdW: a model for supercritical fluids} \label{sec.modelVdW}
         
        The reduced van der Waals equation of state is used and reads
        \begin{equation} \label{eq.vdw}
            \check{p} = \frac{8 \rho \check{T}}{3 - \check{\rho}} - 3 \check{\rho}^2,
        \end{equation} 
        \noindent where the reduced variables $\check{p} = p^* / p_c^*$, $\check{\rho}= \rho^* / \rho_c^*$ and $\check{T}= T^* / T_c^*$ have been introduced, with $p_c^*$, $\rho_c^*$ and $T_c^*$ being the critical pressure, density and temperature, respectively.
        A choice of diffusion laws is required in addition to this equation of state.
        The models proposed by \cite{jossi1962viscosity} and \cite{stiel1964thermal} are used for the dynamic viscosity and the thermal conductivity, respectively.
        They provide analytical expressions for non-polar supercritical fluids based on theoretical scalings and experimental fittings. 
        In supercritical fluids, these diffusion laws depend both on $\check{T}$ and $\check{\rho}$.
        The density, dynamic and kinematic viscosity profiles are plotted in figure \ref{fig.Fluids}.
        Note that the reference temperature, here again noted $T_m^*$ to maintain consistency with the previous fluids, is usually termed pseudo-boiling or pseudo-critical temperature in supercritical fluids \citep{banuti2015crossing}. 
        Density and dynamic viscosity are strongly correlated, and both exhibit strong gradients in the pseudo-boiling region.
        This motivated the choice of fluid HT, where density and viscosity are both defined using a hyperbolic tangent function, aiming at capturing these gradients while neglecting other variations away from them.
        The kinematic viscosity admits a minimum around the pseudo-critical temperature but is not localised, as opposed to the other fluids.
        The relatively simple model of fluid HT is found to decently reproduce this minimum as a result of the shift $\epsT$ introduced between the hyperbolic tangent laws, but differs away from the point where $\nu$ remains strictly constant in fluid HT.
        
        In analogy to the previous fluids, we would like to extract the characteristic scales $\dTnu$ and $\Anu$ from the kinematic viscosity law.
        However, while $\nu$ does have a minimum in fluid VdW, it is not clear that this minimum is localised over a finite, identified range $\dTnu$.
        Still, it can be observed, after calculation, that $\nu(T)$ admits two inflection points in the vicinity of the pseudo-boiling temperature -- one below and one above this temperature.
        This can be used to define the scale $\dTnu$ as the width between these two inflection points.
        From this, an amplitude $\Anu$ can be naturally defined.
        The procedure is thoroughly described in appendix \ref{app.definition_VdW_dnu_Anu}.
        Finally, note that the reduced pressure is the control parameter of the kinematic viscosity seen as a function of the temperature.
        In other words, $\nu(T)$ is different for each $\check{p}$, and, consequently, so are $\dTnu$ and $\Anu$.

    \subsection{Base flow} \label{sec.baseflow}

        Linear stability analysis requires the knowledge of a base flow, defined as a steady solution of the non-linear Navier-Stokes equations.
        After recasting the non-linear Navier-Stokes equations given the physical assumptions associated with this flow, the equations are numerically solved.
        Plane Couette flow is comprised between two plates and is driven by the upper plate moving at speed $u_1^*$, which is used as the reference velocity scale.
        The streamwise and wall-normal directions are noted with $x$ and $y$, respectively. 
        The flow is assumed to be parallel: the streamwise velocity $u$ does not depend on $x$, and the wall-normal and spanwise components of the velocity $v$ and $w$, respectively, are zero.
        The lower plate is fixed and, given the no-slip conditions, the non-dimensional streamwise velocity at the boundaries verifies $u(0)=0$ and $u(1)=1$.
        The distance $h^*$ between the two plates is used as the reference length scale.
        The lower plate is kept at temperature ${T_0}^*$, chosen as the reference temperature.
        We choose to consider the non-dimensional temperature gradient $\tau$ between the two plates as an input parameter, which in turn sets the temperature of the upper plate.
        The boundary conditions for the temperature are then $T(0)=1$ and $T(1)=1 + \tau$. 
        Under the assumption of a steady flow without pressure gradient -- the flow is driven by the top wall, the non-dimensional Navier-Stokes equations reduce to a system of ordinary differential equations:
        \begin{align}
            ( \bar{\mu} \bar{u}' )' &= 0 \label{eq.momentumCouette} , \\
            ( \bar{\lambda} \bar{T}' )' &= 0 \label{eq.energyCouette} ,
        \end{align}

        \noindent where the superscript $'$ denotes the wall-normal derivative and the overbars identify base-flow variables.
        The inertial terms are zero given the parallel flow assumption, and the problem does not depend on the Reynolds and Prandtl numbers.
        Besides, the temperature is decoupled from the velocity field.
        When $\lambda$ is constant, as it is supposed to be in fluids VB, DB and HT, the temperature profile is readily obtained as $T(y) = 1 + \tau y$.
        As for fluid VdW, equation \eqref{eq.energyCouette} is solved using Newton's method by setting the initial guess as the aforementioned linear profile.
        Once $T$ is obtained, the density profile is also known via the equation of state.
        The velocity profile is finally obtained by integration of equation \eqref{eq.momentumCouette} with the knowledge of the dynamic viscosity profile as a function of $T$ and $\rho$.
        Finally, note that we make the arbitrary choice to locate the extremum of kinematic viscosity at the centre line of the flow, $y=1/2$.
        This is achieved by accordingly setting $T_{m}^*/{T_0}^* = 1 + \tau/2$ under the assumption that the temperature profile is linear -- which is indeed the case for fluids VB, DB, and HT.

    \subsection{Inviscid linear stability theory}

    \subsubsection{Rayleigh equation with density gradients} \label{sec.linear_dynamics}

    Assuming infinitely small, inviscid, two-dimensional perturbations, the linearised Navier-Stokes equations in the low-Mach approximation \citep{rehm1978equations,paolucci1982filtering} can be written:
    \begin{align}\label{eq.LNS_LowMach}
    \Derp{u}{x} + \Derp{v}{y} &= 0  , \\    
    \rhobar \left( \Derp{u}{t} + \ubar \Derp{u}{x} \right) &= - \Derp{p}{x} - \rhobar \Derp{\ubar}{y} v \label{eq.LNS_momX} , \\
    \rhobar \left( \Derp{v}{t} + \ubar \Derp{v}{x} \right) &= - \Derp{p}{y}  \label{eq.LNS_momY} , \\
    \Derp{T}{t} +  \ubar \Derp{T}{x}  &=  - \Derp{\Tbar}{y} v , \label{eq.LNS_energy}
    \end{align}        
    Perturbations of the form $q(x,y,t) = \Re \{ \hat{q}(y) e^{i(\alpha x - \omega t)} \}$ are now considered, with ${q = [u, v, T]}$ being the state vector of the perturbations and  $\Re \{ \}$ the real part.
    These linearised equations can then be recast into the Rayleigh equation governing the linear dynamics of incompressible flows with density gradients (see also \cite{fontane2008stability}):
    \begin{equation}\label{eq.Rayleigh}
        \bar{\rho} ( \bar{u} - c) \left[  \hat{v}'' + \frac{\bar{\rho}'}{\bar{\rho}} \hat{v}' - \alpha^2 \hat{v} \right] - (\bar{\rho} \bar{u}')' \hat{v}  = 0 ,
    \end{equation}
    where $c = \omega / \alpha$ is the complex phase velocity.
    Note that the disturbance temperature does not appear in equation \eqref{eq.Rayleigh} since the linearised mass and momentum equations are decoupled from the energy equation \eqref{eq.LNS_energy}.
    Temperature disturbances are deduced from the hydrodynamic disturbances, which can be calculated independently.
    Thermal effects are however at play in the velocity and density profiles of base flow, whose momentum equations \eqref{eq.LNS_momX} and \eqref{eq.LNS_momY}, and ultimately the Rayleigh equation \eqref{eq.Rayleigh}, depend on.
    A temporal framework is adopted: the wavenumber $\alpha$ is a real parameter while the frequency $\omega$ is a complex number that is to be determined.
    The temporal growth rate is given by its imaginary part, $\omega_i$.
    A positive value corresponds to an inviscid instability.
    The (real) phase velocity $c_\varphi$ of the perturbation is simply $c_r$, the real part of $c$. 

    Equation \eqref{eq.Rayleigh} can be classically solved numerically as an eigenvalue problem.
    The boundary condition $\hat{v} = 0$ is used at the wall.
    A pseudo-spectral method is employed to discretise the system and to obtain the derivative matrices \citep{orszag1971accurate}.
    In order to avoid the singularity at the critical layer for neutral modes, a parabolic complex mapping is used, following \cite{boyd1985complex}.
    This allows the growth rate to be computed even when it reaches small values, while a real mapping would produce spurious numerical oscillations.

    \subsubsection{Vorticity} \label{sec.vorticityTheory}

    In order to interpret some results, it can be useful to consider an alternative formulation of the problem in terms of the disturbance vorticity $\xi = \derp{v}{x} - \derp{u}{y}$.
    For a parallel base flow without pressure gradients, $\xi$ is governed, in the physical space, by the linear equation
    \begin{equation} \label{eq.linear_vorticity}
    \Derp{\xi}{t} + \bar{u} \Derp{\xi}{x} = \underbrace{- \Omega' v }_{S_\xi}  \underbrace{ - \frac{\bar{\rho}'}{\bar{\rho}^2}  \Derp{p}{x}}_{B_\xi} ,
    \end{equation}
    where $\Omega$ is the vorticity of the base flow.
    The left-hand side represents the material derivative of $\xi$ by the base flow.
    The right-hand side corresponds to vorticity sources, which may induce an instability.
    The term $S_\xi$ is the production of vorticity responsible for shear flow instabilities.
    The second term, $B_\xi$, is the inertial baroclinic torque, which may generate vorticity when the density and pressure gradients are not aligned.
    In the absence of density gradients, this term is evidently zero.


\section{Criterion of instability based on the kinematic viscosity profile} \label{sec.CouetteGIP}

    A necessary condition for an inviscid instability to exist was given by \cite{rayleigh1880stability} for constant-density flows.
    It requires the existence of an inflection point in the velocity profile of the base flow $\bar{u}''= 0$.
    In the presence of a density gradient, a generalisation of Rayleigh's theorem can be derived, often called the generalised inflection point (GIP) criterion in non-zero Mach number flow studies \citep{lees1946investigation,mack1984boundary}.
    Introducing the quantity ${\Phi = - \bar{\rho} \bar{u}'}$,
    a necessary condition of inviscid instability is that $\Phi ' = 0$
    somewhere in the base flow profile.
    The location where this condition is verified is termed GIP.
    Assuming that a GIP exists, an additional, more restrictive necessary condition of instability was given by \cite{fjortoft1950application}.
    This criterion can be generalised to varying-density flows, stating that a region where
    \begin{equation}{\label{eq.fjortoft_extended}}
        \Phi' (\bar{u} - \bar{u}_s) > 0 ,
    \end{equation}

    \noindent with $\bar{u}_s$ the velocity at the GIP, is required in the base flow profile in order to observe an inviscid instability.
    The proofs of these two results, stated in the case where density gradients are non-zero, straightforwardly follow those given in \cite{schmid2001} for constant density flows by considering equation \eqref{eq.Rayleigh}.
    For a monotonic velocity profile such as that of plane Couette flow, equation \eqref{eq.fjortoft_extended} must be verified everywhere (except at locations where $\Phi' (\bar{u} - \bar{u}_s) = 0$).
    In this case, it is shown in appendix \ref{app.proofMaxPhi} that the generalised Fj\o rtoft's criterion \eqref{eq.fjortoft_extended} is equivalent to observing a maximum of $|\Phi|$ in the base flow profile.
    This extends the well-known interpretation of a maximum of absolute vorticity in constant-density flows.
    Indeed, noting that, under the parallel flow assumption, the vorticity of the base flow $\Omega$ is simply $\Omega = -\bar{u}'$, the quantity $\Phi$ can be interpreted as the density-weighted vorticity:
    \begin{equation} \label{eq.Phi_definition}
        \Phi = \bar{\rho} \Omega.
    \end{equation}
    For constant-density flows, the usual interpretation of the Fj\o rtoft's criterion is then recovered, since $|\Phi| =|\Omega|$ in this case.
    However, in the presence of density variations, a maximum of vorticity is no longer a necessary condition of instability, and the existence of a maximum of $|\Phi|$ should instead be examined.
    Combining $\rho = \mu / \nu$ and the streamwise momentum equation \eqref{eq.momentumCouette}, it follows that
    \begin{equation}
        (\bar{\nu} \Phi)' = 0,
    \end{equation}
    which, after distributing the wall-normal derivative, can be recast as
    \begin{equation} \label{eq.PhiCondition_Couette}
        \frac{\Phi'}{\Phi} = - \frac{\bar{\nu}'}{\bar{\nu}}.
    \end{equation}
    The important result follows: in stratified plane Couette flow, the existence of a maximum of $|\Phi|$ is equivalent to the existence of a minimum of $\bar{\nu}$.
    Because of the generalised Fj\o rtoft's criterion, a minimum of kinematic viscosity in the base flow profile is then a necessary condition of inviscid instability.
    This motivated the choice of the fluid models considered in this paper (section \ref{sec.fluid_models}), which all feature a minimum of $\nu$ and, therefore, potentially exhibit an instability.  
    Finally, note that equation \eqref{eq.PhiCondition_Couette} is specific to plane Couette flow.
    Different criteria may be expected for other shear flows, as discussed in appendix \ref{sec.GIPothers}.


\section{Base flows of fluids VB, DB and HT} \label{sec.baseflowsModel}

\begin{figure}
        \centering
        \includegraphics[angle=-0,trim=0 0 0 0, clip,width=0.99\textwidth]{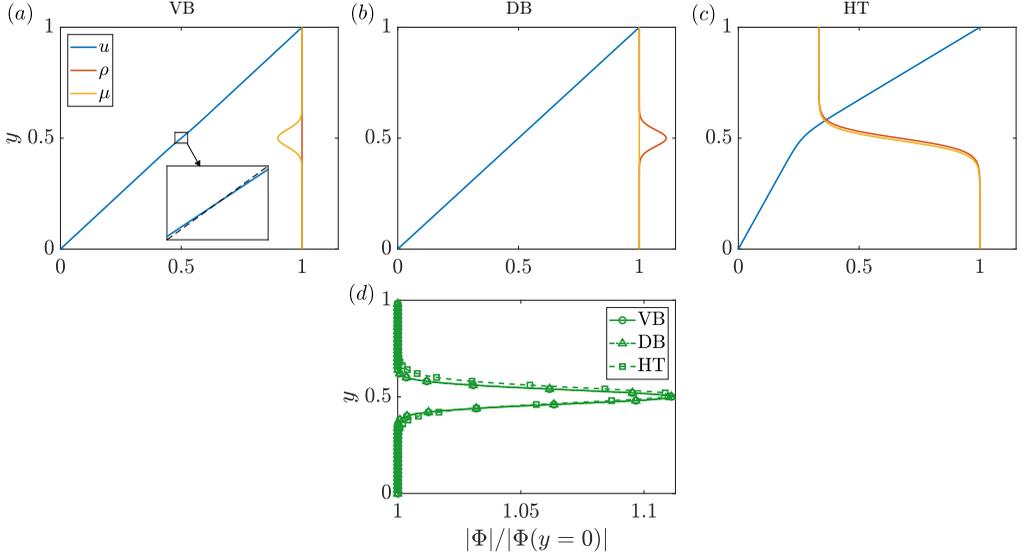}      
        \captionsetup{justification=justified}        
        \caption{Top: base flow profiles of fluids VB, DB and HT, for $\dTnu=10^{-2}$, $\Anu=-10^{-1}$ and $\tau=0.2$. 
        The inset in (a) is a close-up of the velocity profile in the central region, with a comparison to the linear function $f(y)=y$ shown in black dashed line.
        Bottom: resulting profile of $|\Phi|$ for each fluid.}
        \label{fig.baseflow_models}
    \end{figure}

    The base flows associated with the three fluid models VB, DB, and HT are presented in figure \ref{fig.baseflow_models}(a,b,c).
    The density and dynamic viscosity profiles have the same behaviour as those presented in section \ref{sec.fluid_models}  -- the constant temperature gradient of the base flow (section \ref{sec.baseflow}) providing a linear mapping from $T$ to $y$.
    Different velocity profiles are observed. 
    In fluid VB, stronger gradients are present in the centre, where dynamic viscosity decreases. 
    This is a result of the conservation of ${\bar{\mu} \ubar'}$ across the flow, yielding $\ubar' \propto 1/\bar{\mu}$.
    While almost imperceptible in figure \ref{fig.baseflow_models}(a), the presence of these stronger gradients are clearly visible in figure  \ref{fig.baseflow_models}(d), where the profile of $|\Phi|$, as defined in equation \eqref{eq.Phi_definition}, is shown.
    Indeed, in the case of fluid VB, density is constant and $|\Phi|$ reduces $|\Omega|$.
    As for fluid DB, the velocity profile is linear (figure \ref{fig.baseflow_models}b) since viscosity is constant.
    Vorticity is therefore constant, but $|\Phi|$ still assumes a maximum at the centre as it now follows the density profile.
    Turning to fluid HT, the velocity profile features two regions of distinct gradient, which are, again, a consequence of the viscosity distribution (figure \ref{fig.baseflow_models}c).
    The resulting profile $|\Phi|$, exhibiting a maximum in the central region as in the two previous fluids, is here a combination of the variations of density and vorticity.
    In summary, all fluids feature an excess of $|\Phi|$ in the central region.
    This is more generally understood because of the presence of a minimum of kinematic viscosity in each fluid as the integration of equation \eqref{eq.PhiCondition_Couette} leads to $|\Phi| \propto 1 / \bar{\nu}$.
    The normalised profiles of $|\Phi|$ all collapse (figure \ref{fig.baseflow_models}d) since identical parameters $\Anu$ and $\dTnu$ are chosen for each fluid.
    Note that we will only consider $\Anu<0$ in order to generate a maximum of $|\Phi|$, since no instability can occur otherwise according to the generalised Fj\o rtoft's criterion.

    \section{Stability models} \label{sec.analyticalmodel}

        \subsection{Piecewise linear base flows}    \label{sec.piecewiseBF}

    Piecewise linear base flows have been extensively used to study a variety of stability problems with constant density as well as variable density -- usually in the framework of the Taylor-Goldstein equation, under the Boussinesq approximation \citep{drazin1966hydrodynamic}.
    While being simple approximations, useful analytical predictions can be derived from these models, especially predicting the linear stability of long waves \citep{gallaire2015effect}.
    We will here consider arbitrary large variations of density.
    
\begin{figure}
        \centering
        \includegraphics[angle=-0,trim=0 0 0 0, clip,width=0.9\textwidth]{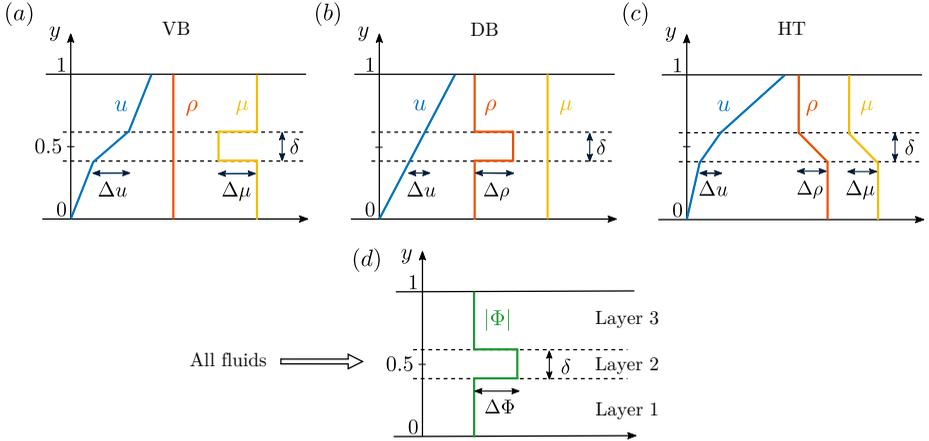}
        \captionsetup{justification=justified}                
        \caption{Piecewise linear base flow profiles for fluids VB, DB, and HT. The central layer (layer 2) is centred around $y=1/2$.}
        \label{fig.RayleighBaseflow}
    \end{figure}

    The base flows of fluids VB, DB, and HT are divided into three layers. 
    The central layer, centred around $y=1/2$, has a width $\delta$ (figure \ref{fig.RayleighBaseflow}).
    This approach follows that proposed by \cite{rayleigh1887stability} for bounded, constant-density flows. 
    Velocity profiles are continuous at the interfaces between layers, but their gradient may not be; a vorticity jump may occur at the interface.
    This is the case of fluid VB, where the viscosity bump is modelled by a discontinuous jump in the central layer (figure \ref{fig.RayleighBaseflow}a).
    This generates a stronger shear rate (i.e., vorticity) in the central layer.
    Note that this is precisely the configuration studied by \cite{rayleigh1887stability}.
    In fluid DB, the shear rate is constant throughout the flow, but density exhibits a jump in the central layer (figure \ref{fig.RayleighBaseflow}b).
    In fluid HT, density linearly varies in the central layer but remains constant in the two other ones (figure \ref{fig.RayleighBaseflow}c).
    The same profile is used for the dynamic viscosity.
    All fluids feature an excess $\dPhi>0$ of $|\Phi|$ in the central layer in order to model the smooth profile of $|\Phi|$ that was observed in the previous section (figure \ref{fig.RayleighBaseflow}d).

        \subsection{Expressions of $\delta$ and $\dPhi$} \label{sec.delta_dphi_definition}
    
    The relations between the parameters $\delta$ and $\dPhi$ of the sought stability model and the physical input parameters of our system $\Anu$, $\dTnu$, and $\tau$ are now examined. 
    The region of the smooth base flow profiles across which $\Phi$ varies is the same as that across which kinematic viscosity varies, as expressed in equation \eqref{eq.PhiCondition_Couette}.
    The characteristic length of this region in the base flow profile is proportional to $\dTnu$, which is related to the fluid property, and inversely proportional to the temperature gradient $\tau$ of the flow.
    The thickness of the layer $\delta$ thus follows the proportionality relation:
    \begin{equation} \label{eq.delta_definition} 
        \delta \propto (1 + \tau/2) \frac{\delta T_\nu}{\tau} ,
    \end{equation}
    where the factor $(1 + \tau/2)$ results from the factor $T_m^* / T_0^*$ (see section \ref{sec.baseflow}) that appears when $\delta T_\nu$ is made dimensionless with $T_0^*$.
    A choice of a prefactor is ultimately required in order to assign a definitive value to $\delta$ in equation \eqref{eq.delta_definition}, and will be specified for each fluid.
    
    We define the quantity $\dPhi$ as the jump of $|\Phi|$ at the interface: $\dPhi = \Delta ( |- \bar{\rho} \bar{u}' | )$.
    The following convention is used: $\dPhi > 0$ corresponds to configurations in which the magnitude of $|\Phi|$ is larger in the central layer than in the other layers.
    Given that $\bar{\rho} \bar{u}' = \bar{\mu} \bar{u}' / \bar{\nu}$ and that integrating the momentum equation \eqref{eq.momentumCouette} yields $\bar{\mu} \bar{u}' = K$, with $K$ a positive constant, we can express $\Delta \Phi$ as
    \begin{equation}  \label{eq.jumpPhi_generalExpression}
        \Delta \Phi = K \Delta \left( \frac{1}{\bar{\nu}} \right) .
    \end{equation}
    The characteristic value of $\bar{\nu}$ being $1+\Anu$ in the central region and $1$ elsewhere, the jump of $1/\bar{\nu}$ at the interface reads
    \begin{equation} 
        \Delta \left( \frac{1}{\bar{\nu}} \right) = - \frac{A_\nu}{1 + A_\nu} .
    \end{equation}    
    An excess of $|\Phi|$ in the central layer ($\Delta \Phi > 0$) is associated with a minimum of $\nu$ ($\Anu < 0$), consistent with section \ref{sec.CouetteGIP}.
    The derivation of the different expressions of the constant $K$ and the final expression of $\dPhi$ associated with each fluid model is detailed in appendix \ref{app.K}.

    \subsection{Stability calculations in the long wave approximation} \label{sec.analytical_LongWave}

        \subsubsection{Derivation} \label{sec.methodLongwave}

    The Rayleigh equation \eqref{eq.Rayleigh} is solved for the three piecewise linear base flows introduced in section \ref{sec.piecewiseBF}.
    We will restrict our analysis to long waves ($\alpha \ll 1$).
    Following \cite{rayleigh1887stability} (see also \cite{drazin2004hydrodynamic,charru2011hydrodynamic}), equation \eqref{eq.Rayleigh} is first solved separately in each of the three layers of the piecewise linear base flows.
    As $\Phi$ is constant in each layer, the last term of the Rayleigh equation vanishes.
    Furthermore, assuming $\alpha \ll 1$ and writing $\hat{v}$ and $c$ as a power series of $\alpha$, equation \eqref{eq.Rayleigh} reduces, at the order $\alpha^0$, to
    \begin{equation} \label{eq.Rayleigh_LW}
        ( \rhobar \hat{v}') '  = 0 .
    \end{equation}
    This equation can be solved in each layer.
    When density is constant across a layer, the solution is simply
    \begin{equation} \label{eq.solutionLayer_Rhocte}
        \hat{v}_i = A_{i} y + B_{i},
    \end{equation}
    where the index 'i' refers to the layer 1, 2 or 3 (see figure \ref{fig.RayleighBaseflow}).
    In the central layer of fluid HT, where density varies, only the first derivative of $\hat{v}$ will be needed.
    This is because interfaces conditions, described hereafter, set the value of $\hat{v}_2$ using $\hat{v}_1$ and $\hat{v}_3$ (readily obtained from equation \eqref{eq.solutionLayer_Rhocte}).
    The following expression is immediately found:
    \begin{equation} \label{eq.solutionLayer_Rhovary}
        \hat{v}_2' = \frac{ A_{2} y }{\rhobar} .
    \end{equation}
    At each interface between the layers, the kinematic and dynamic conditions \citep{charru2011hydrodynamic} read
    \begin{align} 
        \Delta \left[ \frac{\hat{v}}{\ubar - c} \right] &= 0   \label{eq.kinematic} , \\ 
        \Delta \left[ (\ubar - c ) \hat{v}' + \Phi \hat{v} \right] &= 0 \label{eq.dynamic}.
    \end{align}
    Note that equation \eqref{eq.kinematic} reduces to $\Delta \hat{v} = 0$ as $\ubar$ is continuous. 
    Using equations \eqref{eq.kinematic} and \eqref{eq.dynamic} as well as the boundary conditions $\hat{v}=0$ at the walls lead to a linear system on the coefficients $A_i$ and $B_i$.
    Equating the determinant to zero provides an expression of $c^2$.
    The system is unstable for $c^2<0$ and stable for $c^2>0$.
    In this section, derivations are carried out in the frame of reference moving with $\ubar(y=1/2) = \ubar_{1/2}$. 
    
        \subsubsection{Results for fluid VB}

    As previously mentioned, the dispersion relation for fluid VB corresponds to that derived by \cite{rayleigh1887stability}.
    We will use his result in the long wave regime. 
    Using geometrical reasoning on the piecewise linear base flow of fluid VB (figure \ref{fig.RayleighBaseflow}), $\Delta u$ can be written as         $\du = \delta \left[ 1 + \dPhi ( 1 - \delta) \right]$, recalling that $\Delta \Phi = \Delta \Omega$ for this fluid. 
    Injecting this into Rayleigh's result yields
    \begin{equation} \label{eq.c2_VB_final}
        c^2 = \frac{\delta}{4} \left( \delta -  \dPhi (1 - \delta)^2 \right).
    \end{equation}
    Note that $\dPhi$ was defined as $\dPhi = \Delta (|\Phi|)$ in section \ref{sec.delta_dphi_definition}, and that $\dPhi>0$ corresponds to an excess of $|\Phi|$ in the central layer.
    The instability criterion is $\dPhi > \dPhi_c$ is deduced, with the instability threshold $\dPhi_c$:
    \begin{equation} \label{eq.criteron_VB}
        \dPhi_c = \frac{\delta}{(1 - \delta)^2} ,
    \end{equation}

        \subsubsection{Results for fluid DB} 

    The expression for $c^2$ in fluid DB is solved by first remarking that $\Delta u = \delta$, given that $\ubar(y) = y$ in this case.
    It is also noticed that $\drho = \dPhi$.
    Analytical calculations lead to the relation
    \begin{equation} \label{eq.c2_DB_final}
        c^2 = \frac{ \delta \left[ \delta - \dPhi (1 - \delta) \right] }{4 \left[ 1 + \dPhi( 1 - \delta) \right] } .
    \end{equation}
    Because $\delta < 1$ and $\dPhi>-1$ (since $\rhobar>0$), the denominator is always strictly positive.
    The instability criterion is thus deduced from the sign of the numerator, leading to $\dPhi > \dPhi_c$, with
    \begin{equation} \label{eq.criteron_DB}
        \dPhi_c = \frac{\delta}{(1 - \delta)}.
    \end{equation}
    
        \subsubsection{Results for fluid HT} \label{sec.analyticalmodelHT}
  
    In fluid HT, the transformation of $\drho$ into the parameters $\delta$ and $\dPhi$ cannot be found.
    We will then use $\drho = 2 \gamma / (1+\gamma)$, which comes from the definition ${\drho = \rhobar(0) - \rhobar(1)}$ introduced in figure \ref{fig.RayleighBaseflow} while using $\rhobar^*(0)$ as the reference scale. 
    The stability model will then depend on $\gamma$ in addition to $\delta$ and $\dPhi$, but will become independent of $\gamma$ in some regimes of interest.
    The approximation $\du = \delta$ will be used, which is valid when the central layer or the viscosity jump are small.
    This will be shown to be of practical interest for the more realistic fluid VdW. 
    Under these considerations, the following expression can be derived:
    \begin{equation} \label{eq.c2_HT_final}
        c^2 = \frac{1}{4} \left\{ \sqrt{ \delta \left[ 1 - \dPhiz (1 - \delta) \right] \left[ \delta - \dPhiz(1-\delta) + \gamma^2   \dPhiz (1 - \delta)^2 \right] }  + \delta \gamma \dPhiz (1 - \delta) \right\}^2 ,
    \end{equation}
    where the quantity
    \begin{equation} \label{eq.dPhi0}
        \dPhiz = \frac{ \dPhi }{1 - \gamma} ,
    \end{equation}
    is introduced to simplify equation \eqref{eq.c2_HT_final}.
    It will also be shown to be of practical interest for fluid VdW as $\dPhiz$ only depends on the jump of kinematic viscosity $\Delta (1/\bar{\nu})$ (appendix \ref{app.K}).
    A criterion of instability is obtained by examining the sign of the expression under the square root in equation \eqref{eq.c2_HT_final}.
    Given that the factor ${1 - \dPhiz (1 - \delta)}$ is always positive, the criterion of instability is given by the third factor, reading $\dPhi > \dPhi_c$  with
    \begin{equation} \label{eq.criteron_HT}
        \dPhi_c = \frac{\delta ( 1 - \gamma)}{(1 - \delta) \left[ 1 + \gamma^2 (1 - \delta) \right]}.
    \end{equation}

    \subsubsection{Comments and limiting case} \label{sec.limitingcase}

    The three different criteria all state that a certain excess of $|\Phi|$ is required in the central layer ($\dPhi > \dPhi_c$) to generate an instability.
    This can be seen as an improvement of the Fj\o rtoft's criterion, which states that an instability may occur only if $\dPhi>0$, but does not specify the magnitude of the excess of $|\Phi|$ that is sufficient to make the system unstable.
    It can be noticed that the derived instability thresholds always increase with $\delta$.
    In the limit of small $\gamma$, fluid DB and fluid HT possess the same criterion of instability, equation \eqref{eq.criteron_HT} reducing to equation \eqref{eq.criteron_DB}.
    Furthermore, in this regime, these equations differ from that of fluid VB (equation \eqref{eq.criteron_VB}) only at order $\mathcal{O}(\delta^2)$.
    At order $\mathcal{O}(\delta)$, all three fluids share the common criterion of instability:
    \begin{equation} \label{eq.NC_smallDelta}
        \dPhi > \delta .
    \end{equation}
    Moreover, the growth rate near the instability threshold can be calculated from the different expressions of $c^2$, using $\dPhi = \mathcal{O}(\delta) \ll 1$.
    A general expression is obtained for the unstable modes of all fluids:
    \begin{equation} \label{eq.growhrateSmallDeltaSmalldPhi}
        \omega_i = \alpha \sqrt{ \delta ( \dPhi - \delta ) } ,
    \end{equation}
    This shows the fundamental role the quantities $\dPhi$ and $\delta$ play in modelling these instabilities, regardless of the types of stratifications.

    \subsubsection{Phase velocity} \label{sec.phasevelocity}

    The equations on $c^2$ obtained for each fluid also provide interesting results regarding the phase velocity of the unstable modes.
    For fluids VB and DB, if $c^2 < 0$, then $c$ is purely imaginary.
    Therefore, an unstable mode will have a phase velocity $\ubar_{1/2}$, which, by symmetry, is equal to $1/2$ for these fluids.
    This does not hold for fluid HT, for which the phase velocity is shifted from $\ubar_{1/2}$ by ${\delta \gamma \dPhiz (1 - \delta)}$.
    By integrating the conservation of shear stress in the three layers, the following expression can be obtained for fluid HT:
    \begin{equation}\label{eq.u12_HT}
        \ubar_{1/2} = \frac{1}{2} - \frac{\gamma (1 - \delta)}{2} .
    \end{equation}
    Note that the departure from $1/2$ in equation \eqref{eq.u12_HT} is of order $\gamma$, while the aforementioned additional shift is at most $\gamma^2$ for small $\dPhiz$ given that ${\dPhiz \sim \gamma | \epsT / \dTmurho | < \gamma }$ (appendix \ref{app.K}).
    Equation \eqref{eq.u12_HT} is therefore expected to be a good approximation of the phase velocity.

\section{Numerical stability calculations for fluids VB, DB and HT} \label{sec.LSA_calculations_models}

        \subsection{Growth rate and phase velocity} \label{sec.growthrate}

    Numerical stability calculations are carried out for fluids VB, DB, and HT using three different thicknesses of the central layer $\delta$.
    Note that $\delta$ is initially not an input parameter of the problem: it is calculated from equation \eqref{eq.delta_definition}, which requires a prefactor.
    This prefactor, which is a priori different for each fluid, is set so as to yield the best agreement between the calculated and predicted stability diagram, which will be presented in the next subsection. 
    The value 1.12 is used for fluids VB and DB.    
    While not fully predictive -- this value is not obtained by the model and requires one calculation point in order to be calibrated, it remains close to one: the model can provide order-of-magnitude predictions even without further knowledge.
    A prefactor equal to 1 is used for HT, requiring no external data.

\begin{figure}
        \centering
        \includegraphics[angle=-0,trim=0 0 0 0, clip,width=0.99\textwidth]{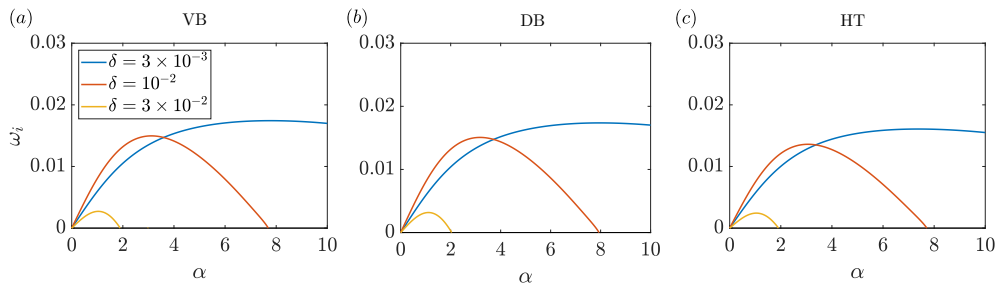}      
        \captionsetup{justification=justified}
        \caption{Growth rate, as a function of $\alpha$, of the fluid models VB, DB, and HT obtained for $\Anu = -0.04$ and different thicknesses of the central layer $\delta$.
        }        
        \label{fig.GR_fluidModels_falpha}
    \end{figure}

\begin{figure}
        \centering
        \includegraphics[angle=-0,trim=0 0 0 0, clip,width=0.99\textwidth]{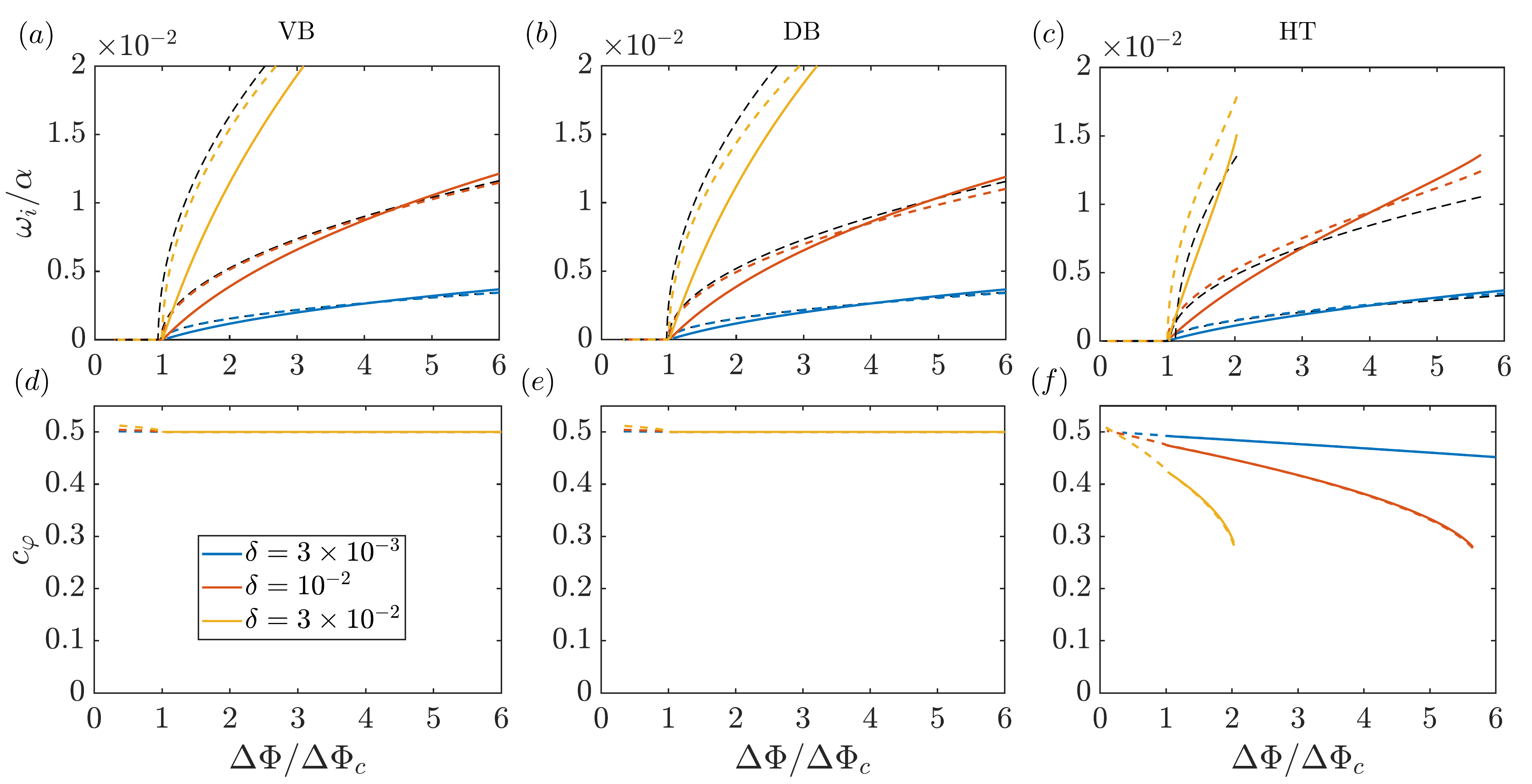}      
        \captionsetup{justification=justified}
        \caption{Top: growth rate, normalised by $\alpha$, of long waves, obtained at $\alpha = 10^{-2}$.
        Bottom: associated phase velocities.
        Solid lines are the numerical results and dashed lines are the theoretical predictions given, for each fluid, in equations \eqref{eq.c2_VB_final}, \eqref{eq.c2_DB_final} and \eqref{eq.c2_HT_final}.
        Black dashed lines correspond to the theoretical general growth rate of equation \eqref{eq.growhrateSmallDeltaSmalldPhi}, obtained for $\delta \ll 1$ near the instability threshold.
        Instability thresholds $\dPhi_c$ are given in equations \eqref{eq.criteron_VB}, \eqref{eq.criteron_DB} and \eqref{eq.criteron_HT}.
        Theoretical phase velocities are given in the frame of reference associated with the lower wall using the values of $\ubar_{1/2}$ discussed in section \ref{sec.phasevelocity}.
        For all figures, each colour is associated with a size $\delta$ of the central layer and each column is associated with a fluid (VB, DB, and HT).
        }        
        \label{fig.GR_fluidModels_compareTheory}
    \end{figure}

    First, a constant value of $\Anu=-0.04$, which sets the magnitude of $\dPhi$ for small $\delta$ \eqref{eq.dPhi_smallDeltaSmalldPhi}, is chosen for all fluids.
    Table \ref{tab.casesGR} indicates the corresponding values of $\dPhi/\dPhi_c$, providing a useful reference when comparing with figure \ref{fig.GR_fluidModels_compareTheory}, which will be presented later.
    As shown in figure \ref{fig.GR_fluidModels_falpha}(a,b,c), similar behaviours as well as close quantitative values of the growth rate are found for all fluids, despite the fundamentally different stratifications in each fluid.
    All fluids exhibit a long-wave instability: low wave numbers are always unstable, while a cutoff wave number $\alpha_c$ exists beyond which the system is stable.
    This observation justifies the restricted analysis to long waves to predict the stability of the system (section \ref{sec.analytical_LongWave}).
    As $\delta$ increases, $\alpha_c$ decreases, as it would for a constant-density, unbounded shear layer for which ${\alpha_c \sim 1/\delta}$ \citep{charru2011hydrodynamic}.
    At constant $\Anu$, the maximum growth rate increases as $\delta$ is reduced.
    However, this does not hold for the growth rate of long waves, for which confinement can have a destabilising effect -- increasing $\delta$ being equivalent to approaching the walls closer to the central layer.    
    This behaviour is not unexpected given that $c^2$ depends on polynomials of $\delta$ of degree larger than 1 (section \ref{sec.analytical_LongWave}).
    Differentiating equation \eqref{eq.growhrateSmallDeltaSmalldPhi} with respect to $\delta$, a general estimate of the value $\delta_m$ that yields the maximum growth rate is found to be ${\delta_m = \dPhi/2}$.
    Using equation \eqref{eq.dPhi_smallDeltaSmalldPhi} for $\dPhi$, we find that $\delta_m \simeq 0.02$, which is consistent with the observations.
    Note that a different behaviour is observed in a bounded, constant-density mixing layer, in which \cite{healey2009destabilizing} found that confinement reduces the temporal growth rate of the instability.

    \begin{table}
     \begin{center}
    \def~{\hphantom{0}}
     \begin{tabular}{lllc}
         Fluids \hspace{0.52cm} & $\delta = 3 \times 10^{-3}$ \hspace{0.8cm}  & $\delta = 10^{-2}$ \hspace{0.6cm}  &   $\delta = 3 \times 10^{-2}$  \hspace{0.2cm}\\[5pt]
         VB  & 13.8  & 4.08 &  1.31  \\
         DB  & 13.8  & 4.13 &  1.35  \\
         HT  & 14.3  & 4.26 &  1.39  \\
     \end{tabular}
     \captionsetup{justification=justified}
     \caption{Values of $\dPhi/\dPhi_c$ for the different fluids and values of $\delta$ presented in figure~\ref{fig.GR_fluidModels_falpha}.}
     \label{tab.casesGR}
     \end{center}
    \end{table}
    
    The growth rate of long waves is presented in figure \ref{fig.GR_fluidModels_compareTheory}(a,b,c).
    The quantity ${\omega_i / \alpha}$ is plotted, which corresponds to the slope of the growth rate at $\alpha=0$.
    For all fluids, the instability threshold is well predicted.  
    The behaviour of the growth rate past the threshold is also reasonably well captured, but piecewise linear models do not yield exact quantitative matches. 
    The validity of the general approximation of the growth rate \eqref{eq.growhrateSmallDeltaSmalldPhi} is verified for $\delta=0.003$.
    At $\delta=0.01$, this approximation still captures the threshold well.
    Yet, significant departures from the full theoretical prediction are observed for fluids DB and HT; for these fluids, more terms are indeed neglected in the derivation leading to equation \eqref{eq.growhrateSmallDeltaSmalldPhi}.
    At larger $\delta$, noticeable differences appear for all fluids.
    Finally, the phase velocity of the unstable mode in each fluid is very well predicted by theoretical models (figure \ref{fig.GR_fluidModels_compareTheory}d,e,f).
    It is equal to $1/2$ for fluid VB and DB, and does not depend on $\delta$ and $\dPhi$.
    The phase velocity markedly differs from $1/2$ in fluid HT as the velocity of the base flow $\ubar_{1/2}$ depends on both $\delta$ and the viscosity ratio $\gamma$ (on which $\dPhi$ depends), as discussed section \ref{sec.phasevelocity}.

        \subsection{Stability diagrams}        

\begin{figure}
        \centering
        \includegraphics[angle=-0,trim=0 0 0 0, clip,width=0.45\textwidth]{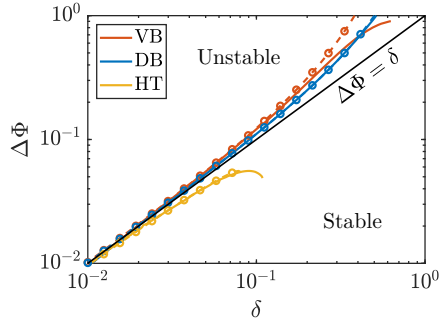}      
        \captionsetup{justification=justified}
        \caption{Neutral curves in the space $(\delta, \dPhi)$, obtained for fluids VB, DB, and HT.
        Solid lines are numerical calculations.
        Dashed lines with circles are theoretical predictions, given in equations \eqref{eq.criteron_VB}, \eqref{eq.criteron_DB} and \eqref{eq.criteron_HT}.}
        \label{fig.collapse}
    \end{figure}

    The stability diagram of each fluid is represented in the space $(\delta,\dPhi)$ in figure \ref{fig.collapse}.
    Neutral curves are calculated by detecting, for each $\dPhi$, the value of $\delta$ for which $\omega_i / \alpha$ becomes close to zero -- the threshold being set to $10^{-4}$.
    Calculations are carried out for long waves at $\alpha=10^{-2}$.        
    Theoretical predictions are generally in good agreement with the numerical results for all fluids.
    For fluid VB, good predictions are observed at low $\delta$, but important discrepancies appear for $\delta>0.2$.
    Such a mismatch is not observed in fluid DB, for which the neutral curve is still accurately predicted for $\delta=0.5$, corresponding to a configuration in which the central layer occupies half of the flow.
    As for fluid HT, the prediction is also excellent.
    Note that, for this fluid, the range of $\dPhi$ that can be studied is limited by the range of $\Anu$ (appendix \ref{app.K}), which is itself limited by $\gamma<1$ and the set value of $\epsT / \dTmurho$ (appendix \ref{app.FluidHT}).
    As predicted in equation \eqref{eq.NC_smallDelta}, the neutral curve of all fluids collapses in the limit of $\delta \ll 1$, following $\dPhi = \delta$.
    As $\delta$ increases, higher-order terms in $\delta$ destabilise fluid HT: the magnitude of $\dPhi$ required to produce an instability becomes smaller than $\delta$.
    Conversely, higher-order terms stabilise fluids VB and DB.

        \subsection{The different sources of vorticity production} \label{sec.vorticityResults}

\begin{figure}
        \centering
        \includegraphics[angle=-0,trim=0 0 0 0, clip,width=0.82\textwidth]{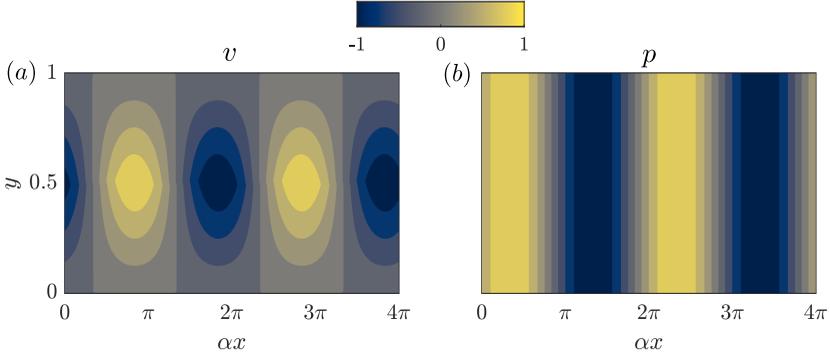}      
        \captionsetup{justification=justified}
        \caption{Wall-normal velocity (a) and pressure (b) of the unstable mode in fluid HT at $\alpha=10^{-2}$, $\delta = 10^{-2}$ and ${\Anu = -0.04}$, plotted in the physical space $(x,y)$ at an arbitrary time $t$.}
        \label{fig.vandP_maps}
    \end{figure}

    The generation of disturbance vorticity can be examined from the structure of the unstable modes in the physical space.
    The wall-normal velocity perturbations are made of a plane progressive wave along~$x$ (figure \ref{fig.vandP_maps}a).
    A peak is observed in the central region of the flow, consistent with the linear increase from zero at the wall predicted in the outer regions in equation \eqref{eq.solutionLayer_Rhocte}.
    The associated pressure field is a plane progressive wave with a phase shift of $\pi/2$ with respect to $v$.    
    Note that results are here presented for fluid HT, but nearly identical fields are obtained for fluids VB and DB.
    
\begin{figure}
        \centering
        \includegraphics[angle=-0,trim=0 0 0 0, clip,width=0.99\textwidth]{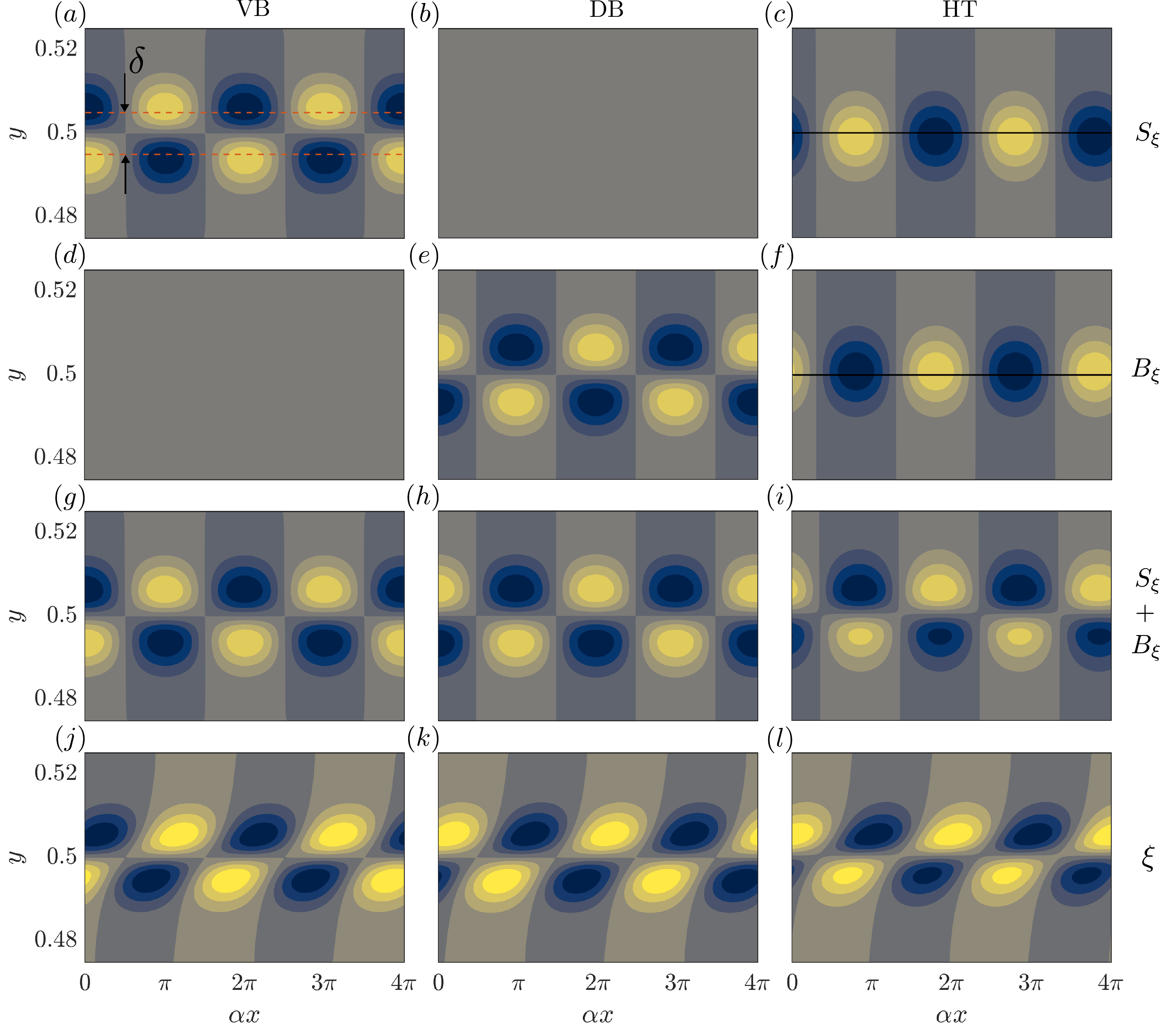}      
        \captionsetup{justification=justified}
        \caption{Terms of the vorticity equation \eqref{eq.linear_vorticity}, plotted in the physical space $(x,y)$ at an arbitrary time $t$, for the unstable modes presented in figure \ref{fig.vandP_maps}.
        The shear term, the baroclinic term, their sum, and the resulting vorticity are shown from top to bottom, in that order.
        Each column corresponds to one fluid.        
        In each figure, fields are normalised by their maximum absolute value, and colour bars are the same as in figure \ref{fig.vandP_maps}.
        The central layer of thickness $\delta=10^{-2}$ is shown in (a) in red dashed line; its location is identical in each figure, albeit not reproduced in order to ease visualisation.
        The location $y=1/2$ is indicated in black solid line in (c) and (f), revealing an offset of the wave below and above this line, respectively.}
        \label{fig.vorticity_maps}
    \end{figure}

    In fluid VB, vorticity production only results from the shear term $S_\xi$ in equation \eqref{eq.linear_vorticity}, given that $\rhobar'=0$.
    Given its mathematical expression, $S_\xi$ follows the same wave structure as $v$, multiplied by the factor $-\Omega'$.
    The vorticity profile $|\Omega|$ of the base flow, which is equal to $|\Phi|$ for this fluid, is only non-zero around the central layer.
    It features strong positive and negative gradients around the lower and upper interfaces, respectively (figure \ref{fig.baseflow_models}d). 
    As a result, the structure of $S_\xi$ contains two out-of-phase waves that are localised around each interface of the central layer, as seen in figure \ref{fig.vorticity_maps}(a).
    The field of total vorticity production, ${S_\xi + B_\xi}$, has the same structure as $S_\xi$ since $B_\xi=0$.

    A similar reasoning can be applied to fluid DB, in which only inertial baroclinic effects are at play given that the base flow vorticity is constant.
    The structure of the inertial baroclinic term $B_\xi$ (equation \eqref{eq.linear_vorticity}) is deduced from that of $p$ and the profile of $- \bar{\rho}'/\bar{\rho}^2$.
    Given that $\rhobar = \Phi$ for this fluid, the $B_\xi$ field presented in figure \ref{fig.vorticity_maps}(e) is found to be similar to that of $S_\xi$ observed for fluid VB (figure \ref{fig.vorticity_maps}a). 
    As $S_\xi=0$, it follows that the total vorticity production ${S_\xi + B_\xi}$ also resembles that of fluid VB (figure \ref{fig.vorticity_maps}g,h).

    As for fluid HT, both $S_\xi$ and $B_\xi$ are active in the generation of disturbance vorticity, and their structure (figure \ref{fig.vorticity_maps}c,f) is markedly different from that of two previous fluids.
    Both feature a peak around the central region.
    This is again a result of the profiles of $- \Omega$ and $- \bar{\rho}'/\bar{\rho}^2$.
    Moreover, $S_\xi$ and $B_\xi$ exhibit a phase difference of $\pi$.
    This is readily understood from the phase difference of $\pi/2$ between $v$ and $p$, another phase shift of $\pi/2$ being added to $B_\xi$ as it contains $\derp{p}{x}$ (equation \eqref{eq.linear_vorticity}).
    Despite being out-of-phase and having a similar structure, the sum of $S_\xi$ and $B_\xi$ is not zero.
    Instead, $S_\xi+B_\xi$ is composed of two out-of-phase waves around each interface (figure \ref{fig.vorticity_maps}i), similar to what was observed for fluids DB and VB.
    This behaviour can be linked to the profile of $\Phi$ with the following arguments.
    Around the central region, the $x$-momentum equation \eqref{eq.LNS_momX} can be approximated to ${{\Phi v \simeq - \derp{p}{x}}}$.
    This results from $\derp{u}{t} + \ubar \derp{u}{x}$ being much smaller than $\Phi v$ in this region, given that the phase velocity of $u$ is here close to $\ubar$, and that the growth rate $\omega_i$ is small (this is more evident in the spectral space, where this term is simply ${i \alpha (\ubar - c) u}$).
    Under this approximation, the linearised vorticity equation \eqref{eq.linear_vorticity} can be recast as
    \begin{equation} \label{eq.vorticity_appproxFinal}
         \Derp{\xi}{t} + \ubar \Derp{\xi}{x} \simeq  - \frac{\Phi'}{\rhobar} v,
    \end{equation}
    where the right-hand side corresponds to $S_\xi+B_\xi$; this shows how the quantity $\Phi$ encapsulates both shear and inertial baroclinic effects.
    The profile of $\Phi'$ and the structure of $v$ ultimately explain the structure of $S_\xi+B_\xi$ in fluid HT (figure \ref{fig.vorticity_maps}i).    
    Note that the denominator ${\rhobar}$, attached to $\Phi'$ in equation \eqref{eq.vorticity_appproxFinal}, modulates the amplitude of the minimum and maximum of $\Phi'$; it can be seen, in figure \ref{fig.vorticity_maps}(i), that larger peaks observed around the upper interface than those around the lower interface, given that $\rhobar$ is smaller in the upper region.

    Overall, the structure of the total vorticity production $S_\xi+B_\xi$ is similar for each fluid, regardless of the fluid stratification.    
    The associated vorticity fields are finally displayed in figure \ref{fig.vorticity_maps}(j,k,l). 
    Their structure follows that of $S_\xi+B_\xi$, with an alteration resulting from advection effects (left-hand side of the vorticity equation \eqref{eq.linear_vorticity}).
    The final picture is two vorticity waves travelling along each interface, with a phase difference of $\pi$ minus a phase shift induced by advection.
    It can also be noted that these waves are generated around the critical layer $y_c$, defined as $\bar{u}(y_c)= c_\varphi$.
    In fluids VB and DB, $c_\varphi=1/2$ (section \ref{sec.phasevelocity}) which, by symmetry of the base flow, leads to $y_c=1/2$.
    In fluid HT, the phase speed of the mode presented in figure \ref{fig.vorticity_maps} is $c_\varphi \simeq 0.45$; we have verified that $\bar{u}(1/2)\simeq 0.45$.  
    
    The relative amplitudes of the terms plotted in figure \ref{fig.vorticity_maps} are hidden by the normalisation of each field.
    The maximum absolute value that each field reaches within the physical space $(x,y)$ is used in the normalisation procedure.
    The relative values are indicated in table \ref{tab.amplitudesFigMech}, noting $M_S$, $M_B$ and $M_{B+S}$ the maxima reached by $S_\xi$, $B_\xi$ and $(S_\xi+B_\xi)$, respectively.
    Results are straightforward for fluids VB and DB for which $B_\xi=0$ and $S_\xi=0$, respectively.
    As for fluid HT, it is shown that the shear and baroclinic effects act with similar strengths.
    Furthermore, the total source of vorticity $(S_\xi+B_\xi)$ is about six times weaker than these effects as it proceeds from the interference of the two waves, cancelling out a large part of their amplitude.

    \begin{table}
     \begin{center}
    \def~{\hphantom{0}}
     \begin{tabular}{lll}
         Fluids \hspace{0.75cm} & $M_S/M_{B+S}$ \hspace{0.5cm}  & $M_B/M_{B+S}$ \hspace{0.2cm}\\[5pt]
         VB  & 1    & 0     \\
         DB  & 0    & 1     \\
         HT  & 6.1  & 6.2 \\
     \end{tabular}
     \captionsetup{justification=justified}
     \caption{Relative maximum amplitudes of the fields shown in figure \ref{fig.vorticity_maps}.}
     \label{tab.amplitudesFigMech}
     \end{center}
    \end{table}

    \subsection{Interpretation} \label{sec.interpretation}

    These results can be interpreted within the wave-interaction theory as reviewed by \cite{carpenter2011instability}.
    In this framework, instabilities are seen as a result of vorticity waves developing along two interfaces that are located close enough so that each wave modifies the velocity field of the other.
    The modified velocity field further deforms each interface, which yields additional vorticity production.
    This forms a positive feedback loop in which the two vorticity waves are amplified, constituting an instability.
    In order to achieve amplification, these waves have to be phase-locked, which was indeed observed for the three fluids in figure \ref{fig.vorticity_maps}(j,k,l).
    The appearance of vorticity waves along each interface has been shown to be driven by the structure of $\Phi$ (equation \eqref{eq.vorticity_appproxFinal}). 
    The physical mechanisms giving rise to these vorticity waves are now further examined.

    In fluid VB, the instability arises from an excess of vorticity in a localised layer.
    This is the essential ingredient of the Kelvin-Helmholtz instability, which is well-known and will not be further discussed.
    The mechanism of this instability has indeed been given following either kinematic \citep{batchelor2000} or dynamic \cite{charru2011hydrodynamic} arguments.
    \cite{carpenter2011instability} also examined this instability from a wave-interaction perspective.

    In fluid DB, only inertial baroclinic effects produce vorticity disturbance.
    This originates from misalignments between gradients of density and pressure \citep{soteriou1995effects}.
    More precisely, given the flow assumptions, this misalignment can only be generated via the density stratification of the base flow and the gradient of pressure perturbations in the streamwise direction.
    Physically, two parcels of fluids at different wall-normal locations, having two different densities (that of the base flow), do not have the same streamwise acceleration when submitted to a streamwise perturbation of pressure \citep{reinaud1999numerical}.
    This induces a wall-normal gradient of streamwise velocity, i.e., vorticity.
    In fluid DB, the central region has an excess of density. 
    The above mechanism of vorticity production, therefore, occurs between the lower and central layer, as well as between the upper and central layer.
    This results in the two vorticity waves that have been observed in figure \ref{fig.vorticity_maps}(h).

    \begin{figure}
        \centering
        \includegraphics[angle=-0,trim=0 0 0 0, clip,width=0.99\textwidth]{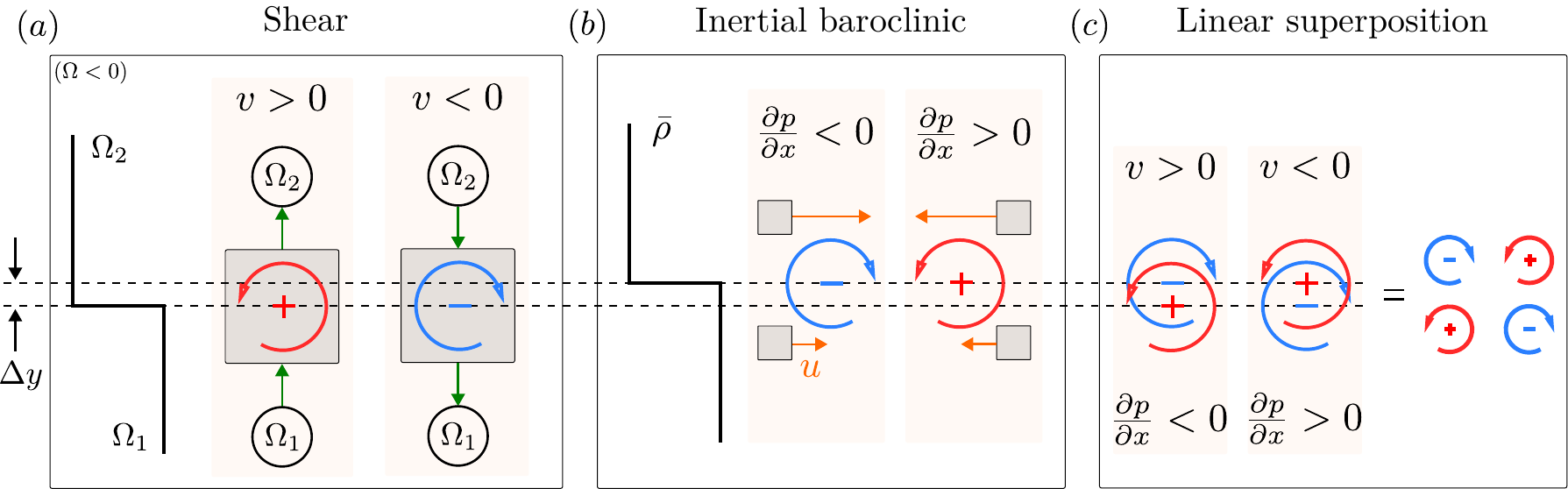}      
        \captionsetup{justification=justified}
        \caption{Sketch illustrating the mechanism generating the two vorticity waves in fluid HT.
        Red and blue vortices indicate positive and negative values of the vorticity disturbance~$\xi$, respectively.        
        These disturbances are generated by shear (a) and inertial baroclinic (b) mechanisms.
        Solid black lines represent the vorticity (a) and density (b) profiles of the base flow, $\Omega$ and $\rhobar$, respectively.
        Note that $\Omega<0$ and that a shift $\Delta y$ exists between the interfaces of $\Omega$ and $\rhobar$.
        Vertical green arrows in (a) represent the transport of $\Omega$ by the disturbance $v$ in and out of a control volume centred around the interface.        
        Horizontal orange arrows in (b) show the disturbance streamwise velocity of a fluid parcel that is induced by a streamwise gradient of disturbance pressure.
        The magnitude of this velocity depends on $\rhobar$, which modifies the inertia of the fluid parcel whether it is located in the lower or upper region.        
        }
        \label{fig.SketchMechanismHT}
    \end{figure}

    As for fluid HT, the generation of the two vorticity waves is not as straightforward. 
    A sketch of the mechanism is shown in figure \ref{fig.SketchMechanismHT}(a,b,c).
    On the one hand, disturbance vorticity is generated following the inertial baroclinic mechanism described in the previous paragraph.
    But contrary to fluid DB, this occurs only between two regions of the flow, the lower and the upper ones, which have different densities (figure \ref{fig.baseflow_models}c) -- the central layer playing the role of an interface between them.
    This idealised representation is illustrated in figure \ref{fig.SketchMechanismHT}(b).
    Therefore, only one interface is felt by the baroclinic perturbations, instead of two as in fluid DB.
    This results in only one vorticity wave, which is generated along the central region, as previously observed in figure \ref{fig.vorticity_maps}(f). 
    On the other hand, disturbance vorticity is also produced by a shear mechanism, which consists in the wall-normal transport of base flow vorticity by the perturbations.
    The base flow can also be divided into two regions of vorticity~-- smaller and larger magnitudes of the shear rate are indeed observed in the lower and upper part of the flow, respectively (figure \ref{fig.baseflow_models}c).
    The central layer again plays the role of an interface between these two regions, and only one interface is felt by the shear perturbations, as sketched in figure \ref{fig.SketchMechanismHT}(a).
    As a result, only one wave is generated by the shear mechanism, as previously observed in figure \ref{fig.vorticity_maps}(c).
    This is a consequence of the plane wave structure of $v$, which, as it takes positive and negative values along $x$, alternatively transports parcels of fluid that contain smaller and larger magnitude of $\Omega$ towards the central region. 
    At this point, each mechanism generates one wave that is localised in the central layer. 
    These two waves are out-of-phase, as discussed in the previous section.
    Moreover, because of the existing shift between the viscosity and density profiles in fluid HT (figure \ref{fig.baseflow_models}c), a small shift $\Delta y \sim \epsT / \tau$ also exists between the interface at play in each mechanism (figure \ref{fig.SketchMechanismHT}).
    As a result, the two central waves are in fact slightly shifted from each other, such that their superposition gives rise to two waves that are localised along each side of the central layer (figure \ref{fig.SketchMechanismHT}c).
    Ultimately, it is these two phase-locked, interacting vorticity waves that produce the instability in fluid HT, following the interpretation of the wave-interaction theory.

\section{Application to supercritical fluids} \label{sec.resultsVdW}

        \subsection{Base flow} \label{sec.baseflowVdW}

\begin{figure}
        \centering            
        \includegraphics[angle=-0,trim=0 0 0 0, clip,width=0.86\textwidth]{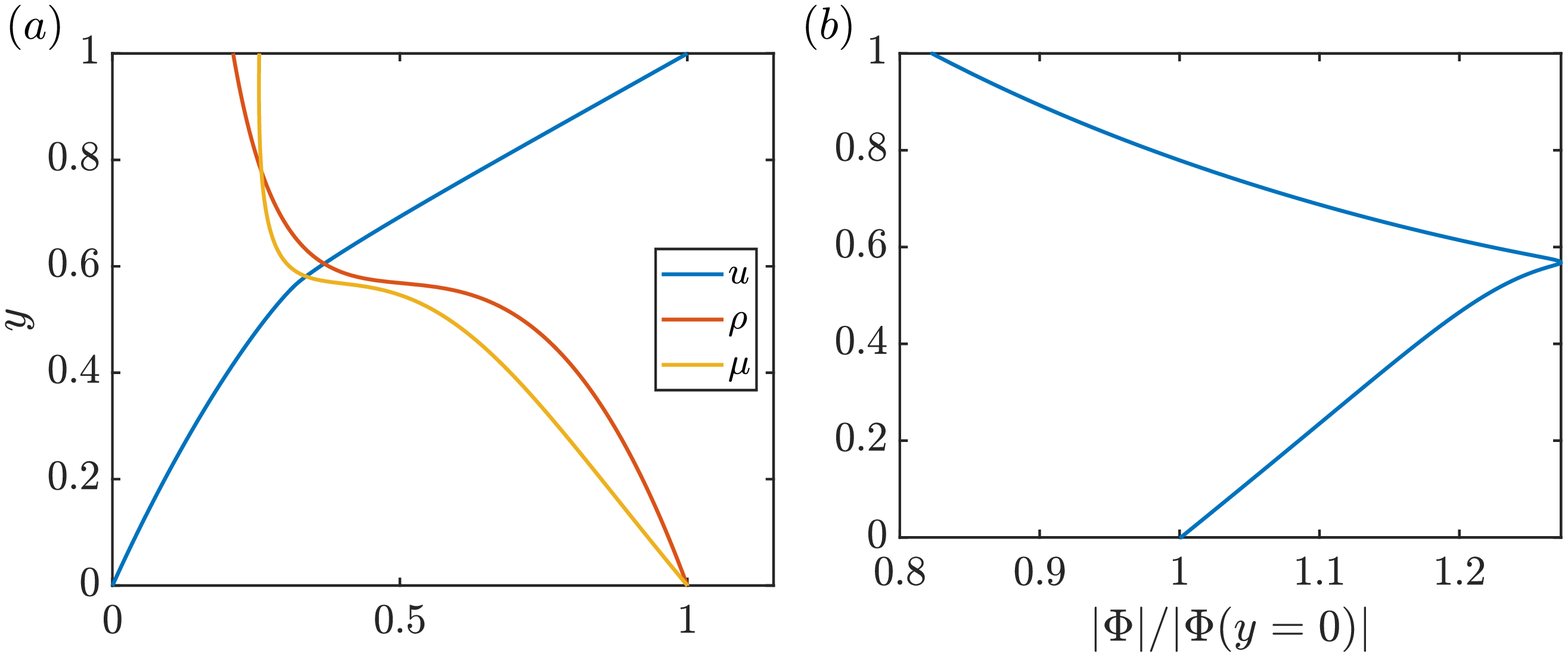}      
        \captionsetup{justification=justified}
        \caption{Base flow profile of fluid VdW at $\check{p}=1.06$ and $\tau=0.5$.}
        \label{fig.baseflowVdW}
    \end{figure}

    A typical base flow of fluid VdW is shown in figure \ref{fig.baseflowVdW}(a).
    The general behaviour is similar to that observed for fluid HT in figure \ref{fig.baseflow_models}(c), the latter being, indeed, an attempt to model some important features of the former.  
    Two regions of markedly different shear rate can be identified.       
    The density and dynamic viscosity profiles are not as simple as in fluid HT: strong gradients are present in the central region, but properties also exhibit weaker variations away from it.
    The resulting profile of $|\Phi|$ \ref{fig.baseflowVdW}(b) follows that of the kinematic viscosity profile, presented in section \ref{sec.modelVdW}.
    As noted, the extremum of $\nu$ (and therefore $|\Phi|$) is seemingly not as localised as in the other fluid models.
    Modelling it with a piecewise view of the $\Phi$ profile, as proposed in figure \ref{fig.RayleighBaseflow}(d), does not appear as an obvious choice.
    However, it was also noted that scales associated with the width and amplitude of the minimum of $\nu$ could be introduced for fluid VdW (appendix \ref{app.definition_VdW_dnu_Anu}).
    By using them, we will show that the piecewise linear model of fluid HT can indeed provide a good representation of fluid VdW, allowing some stability features to be predicted, and thereby providing useful elements to interpret the instability.
    Note that, contrary to the previous fluids, the minimum of $\nu$ is not exactly reached at $y=1/2$ but is shifted upwards.
    This is linked to the procedure used to set the location of the minimum through the reference temperature at the wall and based on the assumption of a constant temperature gradient (see section \ref{sec.baseflow}).
    This assumption is not exact for fluid VdW as thermal conductivity varies, causing the observed shift.

      \subsection{Numerical stability calculations: comparison with the model of fluid HT}
    
\begin{figure}
        \centering            
        \includegraphics[angle=-0,trim=0 0 0 0, clip,width=0.49\textwidth]{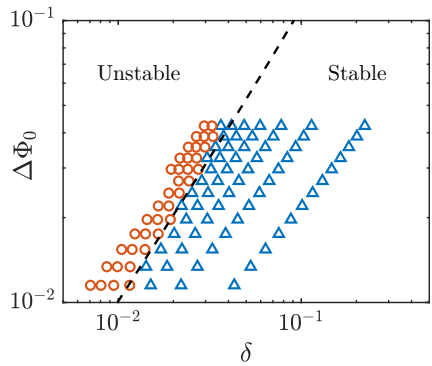}      
        \captionsetup{justification=justified}
        \caption{Stability diagram of fluid VdW for $\check{p} \in [1.03;1.095]$ and $\tau \in [0.005;0.045]$.  
            Red circles and blue squares are numerical results and indicate instability and stability, respectively. 
            The black line is the theoretical prediction given in equation \eqref{eq.NC_VdW}.}
        \label{fig.NC_VdW}
    \end{figure}

    The stability diagram of fluid VdW can be calculated in the space $(\delta, \dPhi)$ by varying both the reduced pressure $\check{p}$ and the temperature gradient $\tau$. 
    Results are compared with the following theoretical predictions obtained for fluid HT in section \ref{sec.analyticalmodelHT}.
    For small $\gamma$, equation \eqref{eq.criteron_HT} leads to the theoretical criterion of instability
    \begin{equation} \label{eq.NC_VdW}
        \dPhiz > \frac{ \delta }{1 - \delta}.
    \end{equation}
    This equation is of interest for fluid VdW as it does not involve $\gamma$ (since $\dPhiz$ only depends on $\Anu$, as shown in appendix \ref{app.K}), which would have had to be defined for such a fluid.
    Note that equation \eqref{eq.NC_VdW} has an error of order $\mathcal{O}(\gamma^2)$ (and not $\mathcal{O}(\gamma)$, as might have been expected), widening the validity of this approximation.
    A generally good agreement is obtained between the numerical calculations and the theoretical prediction (figure \ref{fig.NC_VdW}).
    This shows the robustness of the model based on fluid HT with respect to the property variations outside the central layer and to the upward shift of this layer.
    Furthermore, the definition of $\dTnu$ and $\Anu$ by the inflection point of $\nu(T)$ (appendix \ref{app.definition_VdW_dnu_Anu}) is proved to be relevant, leading to a quantitative prediction of the neutral curve.
    Note that the prefactor used in the definition of $\delta$ (equation \eqref{eq.delta_definition}) is here equal to 1.

\begin{figure}
        \centering            
        \includegraphics[angle=-0,trim=0 0 0 0, clip,width=0.75\textwidth]{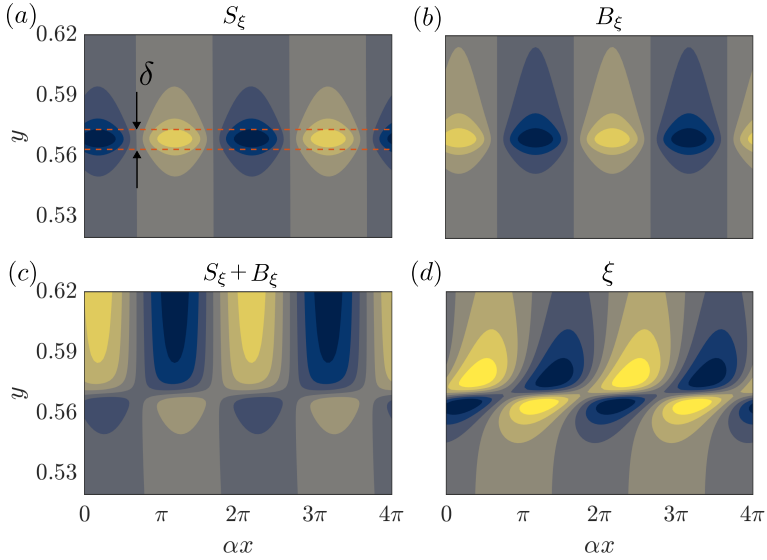}      
        \captionsetup{justification=justified}
        \caption{Terms of the vorticity equation \eqref{eq.linear_vorticity} for the unstable mode in fluid VdW at $\alpha=10^{-2}$, $\check{p}=1.06$ and $\tau=0.5$.
        Normalisation, colour bars, and annotations are identical to those detailed in figure \ref{fig.vorticity_maps}.
        Note that, in fluid VdW, the central region is defined as being centred around the minimum of kinematic viscosity.}
        \label{fig.mechanismVdW}
    \end{figure}

    The different terms of the vorticity equation \eqref{eq.linear_vorticity} for the unstable mode in fluid VdW are now examined.
    The shear term $S_\xi$ (figure \ref{fig.mechanismVdW}a) and the inertial baroclinic term $B_\xi$ (figure \ref{fig.mechanismVdW}b) are each composed of a unique wave, which reaches a maximum in the central region (following the discussion of section \ref{sec.baseflowVdW}, the central region is here defined as being centred around the minimum of kinematic viscosity).
    This is similar to the observations made for fluid HT in figures \ref{fig.vorticity_maps}(c) and \ref{fig.vorticity_maps}(f).
    But because non-zero gradients of $\mubar$ and $\rhobar$ persist away from the central region in fluid VdW (figure \ref{fig.baseflowVdW}a), these terms are not as localised in the centre as in fluid HT.
    This is particularly noticeable for $B_\xi$, which extends further in the upper region, as $\rhobar'$ is non-zero and $\rhobar$ is much smaller than in the lower region.
    The sum $S_\xi+B_\xi$ contains two out-of-phase waves (figure \ref{fig.mechanismVdW}c) given that as $S_\xi$ and $B_\xi$ are themselves out-of-phase and are slightly shifted from each other in the wall-normal direction; see the discussion for fluid HT in \ref{sec.interpretation}.
    However, because of the aforementioned asymmetrical structure of $B_\xi$, the upper wave is not localised around the upper interface of the central layer.
    This constitutes a significant difference with fluid HT (figure \ref{fig.vorticity_maps}i).
    Nevertheless, the final picture is essentially identical: after advection is accounted for, the vorticity field contains two waves localised around each interface, with an additional phase shift leading to a phase difference smaller than $\pi$ between them (figure \ref{fig.mechanismVdW}d).
    The role played by advection into the localisation of $S_\xi + B_\xi$ in the central region can be understood from the vorticity equation \eqref{eq.linear_vorticity}, which can be recast, in the spectral space, as
    \begin{equation}
        |\xi| = \frac{|S_\xi + B_\xi|}{\alpha | \ubar - c |}.
    \end{equation}
    Therefore, $|\xi|$ increases as the phase velocity of the disturbance approaches that of the base flow -- which occurs in the central region, as will be shown in the next subsection.

    Overall, the theoretical stability model developed for fluid HT predicts important features of the stability fluid VdW. 
    This indicates that the main ingredients of the inviscid instability developing in supercritical fluids are indeed contained in fluid HT.
    These ingredients are the presence of strong, localised gradients of density and viscosity and the associated existence of a localised minimum of viscosity -- whose characteristic scales control the instability through the more general parameters $\delta$ and $\dPhi$.  
    The inviscid instability in supercritical fluids can ultimately be interpreted, like in fluid HT, as a result of the combination of shear and inertial baroclinic mechanisms which generate two interacting waves around the central layer (section \ref{sec.interpretation}).

        \subsection{Growth rate and phase velocity}
    
\begin{figure}
        \centering            
        \includegraphics[angle=-0,trim=0 0 0 0, clip,width=0.9\textwidth]{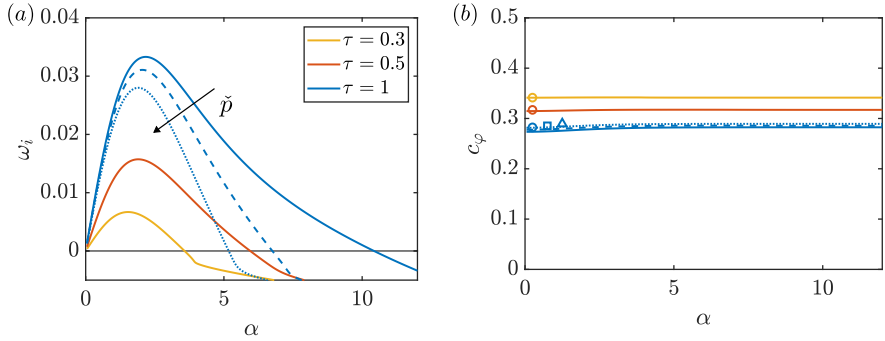}      
        \captionsetup{justification=justified}
        \caption{Growth rate (a) and phase velocity (b) of the unstable mode in fluid VdW. 
        Results in solid lines are obtained for different values of $\tau$ at $\check{p}=1.03$.
        For $\tau=1$, results at $\check{p}=1.06$ and $\check{p}=1.09$ are also presented in dashed line.
        The circles, the square and the triangle in~(b) indicate the value of $\ubar_m$ (see text) at $\check{p}=1.03$, $\check{p}=1.06$ and $\check{p}=1.09$, respectively (these values are not a function of $\alpha$ and are added to this plot for comparison purposes). 
        }
        \label{fig.stability_features_VdW}
    \end{figure}

    Having shown the correspondence between the stability of fluids HT and VdW, further numerical results of fluid VdW can now be examined through the lens of the stability model.
    The growth rate of the unstable mode is shown in figure \ref{fig.stability_features_VdW}(a) for different supercritical pressures ($\check{p}>1$) and temperature gradients $\tau$.
    The long-wave nature of the instability is recovered.
    At constant $\check{p}$, the maximum of the growth rate and the cutoff wavenumber increase with $\tau$. 
    This is consistent with the observations in figure \ref{fig.GR_fluidModels_falpha}(c) for fluid HT, since increasing $\tau$ alone only affects $\delta$, decreasing it.
    Note that the potential destabilisation of long waves by confinement effects (section \ref{sec.growthrate}) is not observed in the present range of parameters.
    When $\tau$ is fixed, the growth rate is reduced as the supercritical pressure increases.
    The interpretation is not straightforward since $\check{p}$ affects both $\Anu$ and $\dTnu$ (appendix \ref{app.definition_VdW_dnu_Anu}), and therefore both $\dPhi$ and $\delta$ in the stability model.
    Equation \eqref{eq.growhrateSmallDeltaSmalldPhi}, which can reasonably be invoked given that the magnitudes of $\dPhi$ and $\delta$ are here of the order of $10^{-2}$, can shed light on this behaviour.
    Both $\dPhi$, which can be approximated by $|\Anu|$ in this regime, and $\delta$, which is proportional to $\dTnu$, increase with $\check{p}$.
    Therefore, the reduction of the growth rate with $\check{p}$ proceeds from a faster decrease of ${(\dPhi - \delta)}$ than the increase of $\delta$.

    Note that negative growth rates are obtained in figure \ref{fig.stability_features_VdW}(a) despite carrying out an inviscid stability analysis, in which only neutral modes are usually expected in the absence of an instability.
    This is a consequence of the use of a complex mapping for $y$ (see section \ref{sec.linear_dynamics}), which is used to remove the singularity of the critical layer for neutral modes.
    Whilst these negative growth rates do not carry any physical significance, this approach allows us to compute accurate cut-off wave numbers, presented later in this section. 

    The phase velocity of the unstable mode is displayed in figure \ref{fig.stability_features_VdW}(b).    
    It is found to be reasonably constant for all wave numbers.
    The velocity of the base flow at the minimum of $\bar{\nu}$, $\ubar_{m}$, provides a good prediction of the phase velocity.
    This is consistent with the results of section \ref{sec.phasevelocity}, substituting $\ubar_{1/2}$ by $\ubar_{m}$ because of the aforementioned upward shift of the central layer.
    This is also consistent with defining the location of the central layer around the minimum of $\bar{\nu}$, as used in the previous subsection.
    A noticeable departure from $\ubar_{m}$ can however be observed as $\tau$ increases, i.e., $\delta$ decreases.
    This behaviour is qualitatively unexpected as the additional shift predicted by the theoretical model should decrease with $\delta$.

\begin{figure}
        \centering            
        \includegraphics[angle=-0,trim=0 0 0 0, clip,width=0.49\textwidth]{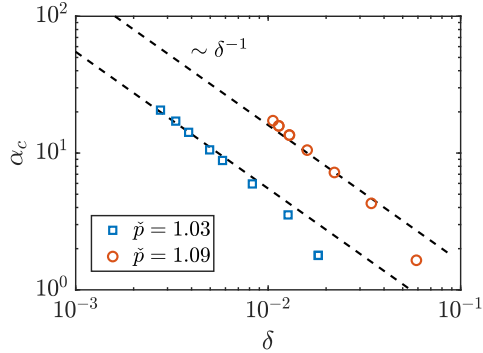}      
        \captionsetup{justification=justified}
        \caption{Cutoff wavenumber obtained for fluid VdW at two reduced pressures.}         
        \label{fig.cutoff_VdW}
    \end{figure}

    The cutoff wavenumber $\alpha_c$ can be plotted from the growth rate calculations by detecting the value of $\alpha \ne 0$ where $\omega_i = 0$.
    Results at two pressures are presented in figure \ref{fig.cutoff_VdW}.
    For each pressure, the cutoff wavenumber is calculated for a range of temperature gradients $\tau$ in order to vary $\delta$. 
    The scaling $\alpha_c \sim \delta^{-1}$ is found.
    While this scaling was not derived from the stability models -- the long wave approximation does not allow $\alpha_c$ to be examined, this result is analogous to that of constant-density mixing layers \citep{charru2011hydrodynamic}.
    Note that the scaling is not valid as $\delta$ increases towards the stability threshold.
    Indeed, as $\delta$ is finite, it would predict finite values of $\alpha_c$ near the threshold, when $\alpha_c$ tends to zero.

\section{Summary and discussion} \label{sec.discussion}

\subsection{Summary}

The stability of strongly density- and viscosity-stratified plane Couette flow was examined.
It was shown that a minimum of kinematic viscosity in the base flow profile produces a GIP, which furthermore satisfies the generalised Fj\o rtoft's criterion.
We first consider three fluid models (termed VB, DB, and HT) which were designed to all exhibit a minimum of $\nu$, while having different stratifications of density and dynamic viscosity.
Modelling their base flow via piecewise linear profiles, the Rayleigh equation that accounts for strong density gradients was solved for each of them in the long-wave approximation.
Expressions of the growth rate and phase velocity were derived, as well as a criterion of instability for each fluid.
All these criteria express that a sufficient excess of $|\Phi|$, the density-weighted vorticity of the base flow, shall be localised in a layer of thickness $\delta$.
This improves, for the specific flow studied in this paper, the generalised Fj\o rtoft's criterion -- which does not provide such an instability threshold.
A general criterion was obtained for all fluids in the limit of small $\delta$, regardless of the type of stratification.

Theoretical predictions were compared to numerical stability calculations.
A qualitatively good agreement was found for the growth rate, while an excellent match was observed for the phase velocity.
Stability diagrams are also generally well predicted by the models, and the collapse of the neutral curves for small $\delta$ was indeed observed in the calculations.
The physical mechanisms responsible for the instability in each fluid were then examined.
For all fluids, two phase-locked vorticity waves travelling along each interface of the central layer were found.
The growth of the instability is ultimately interpreted as a result of the interaction between these two waves.
These waves are generated by either shear effects in fluid VB or inertial baroclinic effects in fluid DB.
In fluid HT, which features strong stratifications of both density and viscosity, the two waves were shown to result from a combination of shear and inertial baroclinic mechanisms.

The stability of a fluid governed by the van der Waals equation of state at supercritical pressure was finally examined.
The stability model was found to quantitatively predict the neutral curve of this more realistic fluid.
The concurrent action of shear and inertial baroclinic effects in the vorticity production was shown to produce two vorticity waves around the central layer.
This suggests that the physical interpretation proposed for fluid HT also provides the essential mechanism of the instability in supercritical fluids.
Ultimately, our study provides evidence that the minimum of kinematic viscosity, reached at the Widom line, is the leading cause of this instability.

\subsection{Supercritical fluids and beyond} \label{sec.discussion_flows}

\begin{figure}
    \centering
    \includegraphics[angle=-0,trim=0 0 0 0, clip,width=0.85\textwidth]{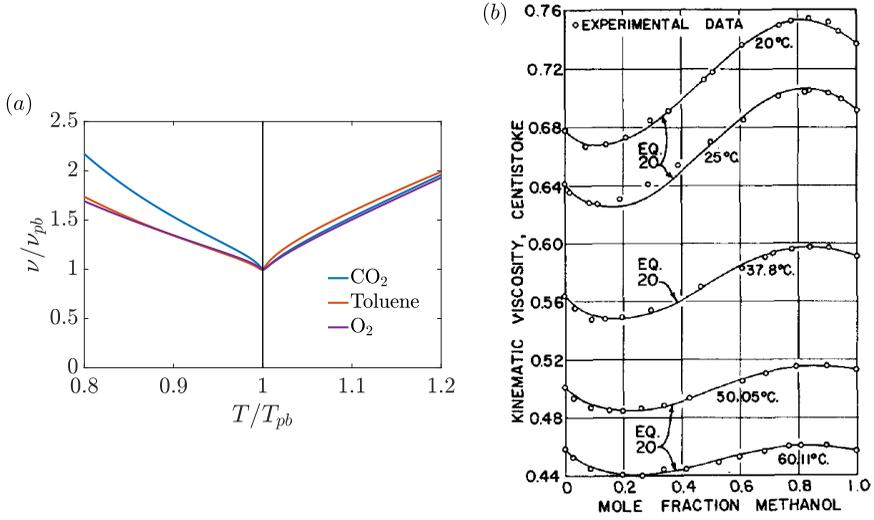}      
    \captionsetup{justification=justified}
    \caption{(a) Kinematic viscosity of different real fluids at supercritical pressure $\check{p}= 1.1$ as a function of temperature.
    The subscript 'pb' refers to the pseudo-boiling point.
    (b)~Kinematic viscosity of a mixture of methanol and toluene as a function of the mole fraction for different temperatures, from \cite{mcallister1960viscosity}.}
    \label{fig.fluids_prop}
\end{figure}

We finally address the link between the present paper and the unstable mode recently found in supercritical fluids by \cite{ren2019boundary}, before discussing how this study may be relevant to other types of fluid and flow.
It has been shown that, in plane Couette flow, a heated supercritical fluid features a similar instability to that developing in simpler fluids, whose only common property was to assume a minimum of kinematic viscosity.
This indicates that other non-ideal thermodynamic effects, such as non-unity compressibility factor $Z$, large values of $c_p$ exhibited near the Widom line or non-monotonic, large variations of the speed of sound, are unlikely to play a decisive role in the instability mechanism.
In the original study of \cite{ren2019boundary}, the existence of the additional unstable mode was reported for a flat-plate boundary layer flow, considering supercritical $\COt$ at non-negligible Mach numbers. 
We aimed at simplifying their configuration by considering a simpler flow (plane Couette), a simpler supercritical fluid model (VdW equation of state and simple analytical diffusion laws) and by neglecting acoustic phenomena.
In this more canonical framework, the instability could be recovered and further analysed.
The role played by the minimum of kinematic viscosity remains to be clearly demonstrated in flat-plate boundary layer flows as studied by \cite{ren2019boundary}, since the theoretical development presented in section \ref{sec.CouetteGIP} cannot directly apply to this flow.
However, the authors observed the additional unstable mode only when the temperature profile crosses the Widom line, about which the presence of a minimum of $\nu$ is a common property (see figure \ref{fig.fluids_prop}a).
This indicates strong links with our results.
\cite{bugeat2022new} furthermore pointed out how, in some cases, inertial effects can be neglected around the pseudo-boiling region, resulting in the correct prediction of a GIP in the boundary layer.
In this case, the momentum equation reduces to that of Couette flow, making the aforementioned theoretical result valid for a flat-plate boundary layer flow.
But a rigorous understanding of the conditions of the existence of a GIP is so far missing in this flow, and further efforts could be directed towards the search for a sufficient condition which would factor in inertial effects. 
The link with an excess layer of $\Phi$ should also be analysed in this case, as the criterion of instability in equation \eqref{eq.NC_VdW}, involving $\dPhi$ and $\delta$, may be altered for a different flow configuration.

Other types of flow may exhibit a localised minimum of viscosity, hence, potentially, a similar instability.
In the present study, the scalar quantity that controls the spatial distribution of $\nu$ is the temperature.
Other quantities could play an analogous role.
The case of a shear flow made of two miscible fluids, mentioned in the introduction, is an interesting example.
The mole fraction is constant in each fluid but varies through the interface, which can be assumed to have a small, finite thickness.
\cite{ern2003stability} studied the instability developing in this system at low Reynolds numbers.
It can be noted that certain fluid mixtures may exhibit a minimum of kinematic viscosity for intermediate mole fraction (between 0 and 1), as shown in figure \ref{fig.fluids_prop}(b).
For these fluids, a minimum of $\nu$ would then exist within the diffused interface, as it does for supercritical fluids in the region of the Widom line.
An inviscid instability can be expected in this case, which may compete with other instability at low Reynolds numbers if viscous damping does not stabilise it too quickly.
As pointed out by \cite{ern2003stability}, a stability analysis of such a system is valid as long as the time scale associated with the diffusion of this interface is large compared to that of the instability.
Similarly, in a single (not necessarily supercritical) fluid flow, heating or cooling a small layer of liquid or gas, respectively, would produce a localised minimum of $\nu$.
An instability could then appear, provided that its associated time scale is small compared to that of thermal diffusion of the cooled or heated layer.
Note that these time scale considerations were avoided in the present study, as the temperature field strictly was a steady solution of the Navier-Stokes equations.
A comparison between the viscous time scale of the perturbations and that of the instability could however evaluate the potential of prediction of the inviscid analysis at finite Reynolds numbers.

Finally, the case of non-Newtonian fluids can be mentioned as dynamic viscosity, and hence kinematic viscosity, may vary as a function of the stress profile.
In a canonical plane Couette, no variation of stress would be observed, and the flow would be linearly stable.
Introducing temperature gradients through viscous heating can alter the stability of the flow \citep{eldabe2007stability}.
In an isothermal configuration, \cite{nouar2009stability} added a streamwise pressure gradient, leading to a plane Couette-Poiseuille flow.
The presence of a maximum (minimum) of stress for a shear-thinning (shear-thickening) fluids may then generate a minimum of kinematic viscosity.
However, the criterion of instability derived in the present paper would not apply because of the presence of a pressure gradient, and an isothermal parallel flow of non-Newtonian fluid may therefore not experience this instability.

\backsection[Declaration of interests]{The authors report no conflict of interest.}

\backsection[Funding]{This work was funded by the European Research Council grant no. ERC-2019-CoG-864660, Critical.}
\appendix

\section{Definition of $\dTnu$ and $\Anu$ in fluid HT} \label{app.FluidHT}
The amplitude of the extremum of kinematic viscosity was previously defined for fluid VB and DB as $\nu_e = 1 + A_\nu$, where $\nu_e$ is the extremum of $\nu$.
Supposing $\nu_e$ is reached around $T=T_m=1$, we have
\begin{equation}
    \nu_e \simeq \frac{\mu_m}{\rho_m} = 1 - \gamma  \tanh \left( - \frac{\varepsilon_T}{\delta T_{\mu,\rho} } \right) .
\end{equation}  

\noindent Assuming that $\varepsilon_T / \delta T_{\mu,\rho}$ is small,
\begin{equation}
    \nu_e \simeq  1 + \frac{\gamma \varepsilon_T}{\delta T_{\mu,\rho}}.
\end{equation}  

\noindent By definition of $A_\nu$, it results
\begin{equation}\label{eq.scalingAnu_tanh2}
    A_\nu =\frac{\gamma \varepsilon_T}{\delta T_{\mu,\rho}}.
\end{equation}

\begin{figure}
    \centering
    \includegraphics[angle=-0,trim=0 0 0 0, clip,width=0.99\textwidth]{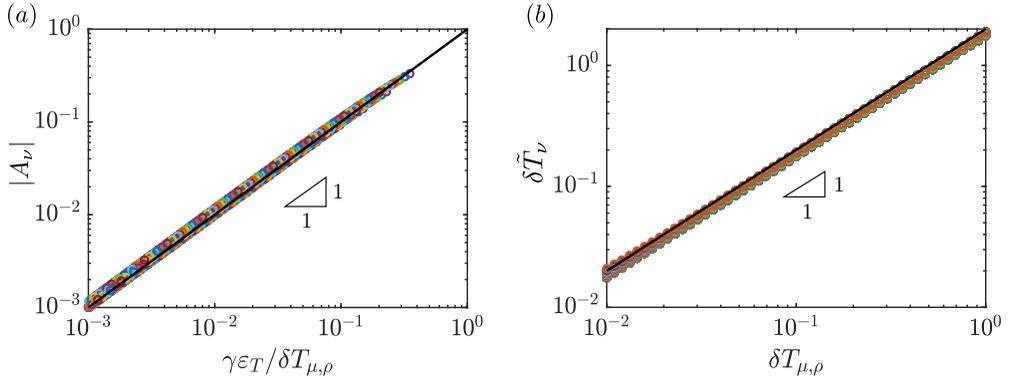}      
    \captionsetup{justification=justified}
    \caption{Verification of the scaling of $\Anu$ (a) and  $\dTnu$ (b) for fluid HT, given in equations \eqref{eq.scalingAnu_tanh2} and \eqref{eq.scalingdTnu_tanh2}, respectively.
    This is achieved by first computing the kinematic viscosity for $\gamma \in [10^{-2};7 \times 10^{-1}]$, $\delta T_{\mu,\rho} \in [10^{-2};1]$ and $\varepsilon_T \in [-5 \times 10^{-3}; -5 \times 10^{-4}]$.
    For each set of parameters, $|\Anu|$ is calculated as $|\nu_e - 1|$, where $\nu_e$ is the minimum of $\nu$. 
    The quantity $\delta \tilde{T}_\nu$ is the full width at half minimum.}
    \label{fig.scalingTanh}
\end{figure}

\noindent This equation is verified in figure \ref{fig.scalingTanh}(a) over a large range of parameters.
It is interesting to note that the direction of the shift between $\rho$ and $\mu$, i.e., the sign of $\varepsilon_T$, determines the nature of the extremum. 

The range of temperature $\dTnu$ containing kinematic viscosity variations must vary as
\begin{equation} \label{eq.scalingdTnu_tanh2}
    \delta T_\nu \propto  \delta T_{\mu,\rho},
\end{equation}
since gradients of $\rho$ and $\mu$ are non-zero only over this characteristic range of temperatures. 
An overlap of the ranges of variation of $\rho$ and $\mu$ is ensured as we assume $\varepsilon_T$ to be small compared to $\delta_T$.        
To check this result, we calculate the quantity $\delta \tilde{T}_\nu$ defined as the width at half maximum or minimum of $\nu$.
This is formally defined as the temperature range where $|\nu - 1| > |\Anu| / 2$. 
The scaling of equation \eqref{eq.scalingdTnu_tanh2} is confirmed in figure \ref{fig.scalingTanh}(b).
An arbitrary choice of prefactor must be made in equation \eqref{eq.scalingdTnu_tanh2} in order to set the definitive expression of $\dTnu$.
We note that a factor of about 2 was observed in figure \ref{fig.scalingTanh}(b), such that $\delta \tilde{T}_\nu \simeq 2 \delta T_{\mu,\rho}$.
Besides, in order to maintain consistency with the Gaussian laws in fluids VB and DB, in which the full width at half minimum is about $2.35 \dTnu / \sqrt{2}$ given the definition of $\dTnu$ in \eqref{eq.model1} and \eqref{eq.model2}, we will ultimately define $\dTnu$ (for fluid HT) as $\dTnu \simeq \delta \tilde{T}_\nu \sqrt{2} / 2.35$.
After round-off, this results in the definition
\begin{equation}
    \dTnu = 1.2 \> \delta T_{\mu,\rho} .
\end{equation}
        
\section{Definition of $\dTnu$ and $\Anu$ in fluid VdW} \label{app.definition_VdW_dnu_Anu}

\begin{figure}
    \includegraphics[angle=-0,trim=0 0 0 0, clip,width=0.97\textwidth]{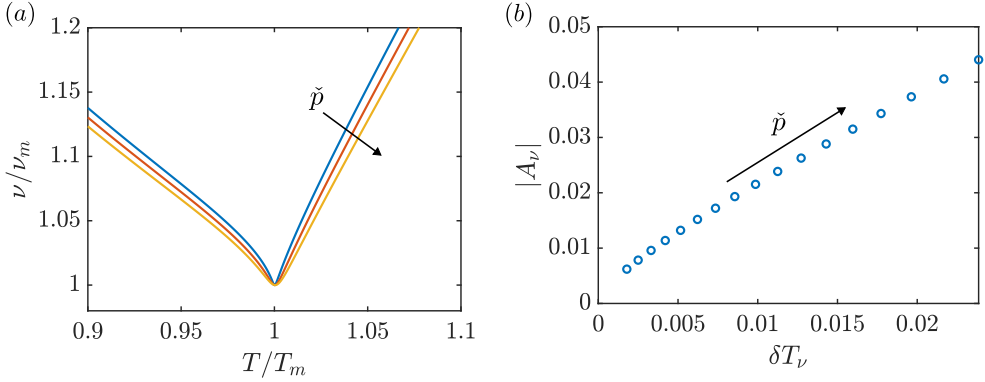}      
    \captionsetup{justification=justified}
    \caption{(a) Kinematic viscosity around the pseudo-boiling point of fluid VdW at supercritical pressures $\check{p}=1.03$, $\check{p}=1.06$ and $\check{p}=1.09$.
    (b) Calculation of the scales $\dTnu$ and $\Anu$ associated with the minimum of $\nu$ in fluid VdW, for $\check{p} \in [1.02;1.10]$.}
    \label{fig.fluidVdW_minimumNu}
\end{figure}

In fluid VdW, the kinematic viscosity profile as a function of temperature depends on the reduced pressure $\check{p}$ (figure \ref{fig.fluidVdW_minimumNu}a). 
For each pressure, a minimum of $\nu$ is reached around the pseudo-boiling point $T_m$.
For a given $\check{p}$, two inflection points of $\nu(T)$ can be calculated.
We note their corresponding temperatures and viscosity $T_1$ and $T_2$ (with $T_1 < T_m < T_2$) and $\nu_1$ and $\nu_2$, respectively.
The scale $\dTnu$ can then simply be defined as 
\begin{equation}
    \dTnu = T_2 - T_1 . 
\end{equation}

\noindent The definition of $\Anu$ is not as straightforward, and we propose the following procedure to extract it.
First, the value of the minimum $\nu_e$, which lies between $T_1$ and $T_2$, is calculated.
Afterwards, we consider $\dTnu$ to be the width at half minimum of $\nu$ as in section \ref{app.FluidHT} for fluid HT.
But contrary to fluid HT, this approach is now used to define $\Anu$ knowing $\dTnu$ rather than the other way round.
Another issue to deal with is the lack of symmetry of $\nu(T)$, resulting in $\nu_1 \neq \nu_2$.
To circumvent this, an average $\nu_{av} = (\nu_1 + \nu_2 )/2$ is defined.
The bump of viscosity is then assumed to be $2( \nu_{av} - \nu_e)$, the factor 2 appearing because of the aforementioned 'half minimum' approach.
Given the definition of $\Anu$ and using $\nu_e + 2( \nu_{av} - \nu_e)$ as the reference viscosity (which can be noted $\nu_\infty$ in analogy with the other fluids), the amplitude $\Anu$ can eventually be defined as
\begin{equation}
    \Anu = \frac{ \nu_e }{ \nu_e + 2( \nu_{av} - \nu_e)}  - 1 ,
\end{equation}
with $\Anu < 0$ indicating that a minimum is reached.
The values of $\dTnu$ and $|\Anu|$ calculated following the above procedure are presented in figure \ref{fig.fluidVdW_minimumNu}(b).
Both decrease as the supercritical pressure $\check{p}$ decreases towards 1, meaning that the minimum of $\nu$ becomes more and more localised as the fluid approaches the critical point.

\section{Proof of the 'maximum of $|\Phi|$' criterion in stratified flows} \label{app.proofMaxPhi}

Let us show that verifying equation \eqref{eq.fjortoft_extended} throughout the base flow, except where ${\Phi' (\bar{u} - \bar{u}_s) = 0}$, leads to the presence of a maximum of $|\Phi|$.
The subscript 's' refers to the variables at the location $y=y_s$, where $\Phi'_s = 0$ as required by the generalised inflection point criterion. 
A Taylor expansion of equation \eqref{eq.fjortoft_extended} can be performed about $y_s$, leading to
\begin{equation} \label{eq.proof1}
    \Phi_s'' (y - y_s) (\bar{u}_s + \bar{u}'_s (y - y_s)  - \bar{u}_s) > 0 .
\end{equation}
Simplifying equation \eqref{eq.proof1} and multiplying by $\bar{\rho}_s$ yields
\begin{equation} \label{eq.proof2}
    - \Phi_s'' \Phi_s (y - y_s)^2 > 0 .
\end{equation}
If $\Phi_s > 0$, equation \eqref{eq.proof2} requires that $\Phi''_s < 0$, meaning that a maximum of $\Phi$ must exist at $y_s$ given that $\Phi'_s =0$.
Doing an analogous reasoning if $\Phi_s <0$, we then conclude that the necessary condition for an inviscid instability to exist, given by the generalised Fj\o rtoft's criterion, is equivalent to the presence of a maximum of $| \Phi |$ at the generalised inflection point.

\section{Notes on the existence of a GIP in pressure-driven flows} \label{sec.GIPothers}

The link between the existence of a GIP that fulfils the generalised Fj\o rtoft's criterion and the existence of a minimum of kinematic viscosity was theoretically investigated in section \ref{sec.CouetteGIP} for plane Couette flow.
This result is not expected to hold a priori in any other shear flows.
The case of the flat-plate boundary layer flow will be discussed later in section \ref{sec.discussion_flows}.
In this section, plane Poiseuille flow is examined in order to point out some major differences.  

Contrary to Couette flow, Poiseuille flow is driven by a constant pressure gradient $G^*$ in the streamwise direction. 
The velocity scale $u_r^* = G^* (h^*)^2/ \mu_0^*$ can be introduced since the flow results from a balance between friction forces and the pressure gradient.
The parameter $h^*$ is the distance between the two fixed plates and $\mu_0^*$ is a reference dynamic viscosity, for example at the lower plate.
The non-dimensional streamwise momentum equation then reads
\begin{equation}\label{eq.Poiseuille_xmom}
   ( \bar{\mu} \bar{u}' )' = 1.
\end{equation}
\noindent Proceeding similarly to section \ref{sec.CouetteGIP}, it follows that
\begin{equation}\label{eq.GIP_PoiseuilleRhoU}
    (\nu \Phi)' = -1. 
\end{equation}  
The right-hand is equal to 1 whereas it was zero in plane Couette flow.
This is due to the pressure gradient driving Poiseuille flow, and this difference will modify the criterion of instability.
To show this, let us first integrate \eqref{eq.GIP_PoiseuilleRhoU}, yielding
\begin{equation}\label{eq.Poiseuille_xmom2}
    \Phi =  \frac{y_0 - y}{\bar{\nu}}, 
\end{equation}
with $y \in [-1/2, 1/2]$, using $h^*$ as the reference length scale.
$y_0$ is the location where $\Phi=0$, i.e., $\bar{u}'=0$.
The existence of such a point inside the domain is guaranteed given the boundary conditions on the velocity profile $\bar{u} (-1/2) = \bar{u} (1/2) = 0$.
Differentiating \eqref{eq.Poiseuille_xmom2} leads to
\begin{equation}\label{eq.GIP_Poiseuille}
    \Phi' = - \frac{1}{\bar{\nu}} \left( 1 + \frac{ (y_0 - y) \bar{\nu}'}{\bar{\nu}} \right).
\end{equation}  
The GIP criterion of instability, $\Phi'=0$, eventually reduces to
\begin{equation}\label{eq.GIP_PoiseuilleFinal}
    \frac{\bar{\nu}'}{\bar{\nu}} = \frac{1}{(y - y_0)},
\end{equation}  
where we ignore the point $y=y_0$ at which no GIP can exist, since $\Phi'(y_0)=1/\bar{\nu}(y_0)>0$.
Besides, because $1 / (y - y_0)$ is strictly non-zero, a minimum of kinematic viscosity is not a condition for the existence of a GIP in plane Poiseuille flow, contrary to plane Couette flow.

Equation \eqref{eq.GIP_PoiseuilleFinal} can be further examined.
Unlike non-heated, constant viscosity Poiseuille flow, the viscosity profile may here break the centre axis symmetry of the velocity profile, which may prevent $y_0$ from necessarily being zero.
As a result, the inequality $|y-y_0|<1$ holds. 
Therefore, the condition \eqref{eq.GIP_PoiseuilleFinal} can only be satisfied if strong enough gradients of viscosity are present in the flow, verifying
\begin{equation}\label{eq.GIP_ConditionPoiseuille}
   \left| \frac{\bar{\nu}'}{\bar{\nu}} \right| > 1 .
\end{equation} 

\noindent We can further write that
\begin{equation}\label{eq.nuy_nut}
   \bar{\nu}' = \left. \Derp{\bar{\nu}}{T} \right|_p \Der{\bar{T}}{y},
\end{equation} 
which, estimating the non-dimensional temperature gradient as $\der{\bar{T}}{y} \sim \Delta T$, gives a necessary condition for the existence of a GIP in plane Poiseuille flow: 
\begin{equation}\label{eq.nut2}
   \left| \frac{1}{\bar{\nu}} \left. \frac{\partial \bar{\nu}}{\partial T} \right|_p \right| > \frac{1}{\Delta T}.
\end{equation}  
This can be interpreted as the need for the kinematic viscosity law of the fluid to contain a temperature scale, defined as $\Delta T_\nu = \nu / (\partial \nu / \partial T )$, that is smaller than the temperature scale $\Delta T$ of the flow.

\section{Expression of $K$ and $\dPhi$ for different fluids} \label{app.K}

    Integrating $\bar{u}' = K / \bar{\mu}$ between 0 and 1 and using the boundary conditions on the streamwise velocity provides an expression of $K$:
    \begin{equation} \label{eq.K}
        K = \frac{1}{\int_0^1 \frac{\mathrm{d} y}{\bar{\mu}}} .
    \end{equation}
    $K$ only depends on the profile of dynamic viscosity.
    
    \subsection{Fluid VB}
    Since $\bar{\rho}$ is constant, the profile of dynamic viscosity is the same as that of kinematic viscosity.
    The integral in equation \eqref{eq.K} can be approximated supposing that $\bar{\mu} = 1+\Anu$ on a layer $\delta$ while $\bar{\mu}=1$ elsewhere.
    This leads to
    \begin{equation} \label{K1}
        K = \frac{1 + A_\nu}{1 + A_\nu (1 - \delta)} .
    \end{equation}    

    \noindent The jump of $|\Phi|$ given by equation \eqref{eq.jumpPhi_generalExpression} can finally be expressed, for fluid VB, as
    \begin{equation} \label{eq.dPhi1}
        \Delta \Phi_{(VB)} = - \frac{A_\nu}{1 + A_\nu ( 1 - \delta)} .
    \end{equation}

    \subsection{Fluid DB}
    Dynamic viscosity is constant, which immediately gives $K=1$.
    The jump of $\Phi$ for fluid DB is therefore
    \begin{equation} \label{eq.dPhi2}
        \Delta \Phi_{(DB)} = - \frac{A_\nu}{1 + A_\nu} .
    \end{equation}

    \subsection{Fluid HT}
    The profile of dynamic viscosity is assumed piecewise linear and equal to the profile of density of fluid HT in figure \ref{fig.RayleighBaseflow}.
    In the lower wall region, $\mubar=1$,
    In the upper wall region, ${\mubar= (1 - \gamma)/ (1+\gamma)}$, after renormalising $\mu^*$ with $\mu_0^*$ in equation \eqref{eq.model3mu}.
    The central region linearly matches these two regions.
    It follows that
    \begin{equation} \label{eq.K_HT}
        \frac{1}{K} = \int_0^{(1-\delta)/2} \mathrm{d}y + \int_0^\delta \frac{ \mathrm{d} \eta }{1 - \frac{2 \gamma \eta}{\delta ( 1 + \gamma)}  } + \int_{(1+\delta)/2}^1 \frac{ \mathrm{d}y }{\frac{1-\gamma}{1+\gamma} }
    \end{equation}   
    where the change of variable $\eta = y - (1+\delta)/2$ was performed for the second integral.
    After integration, we find
    \begin{equation} \label{eq.K_HT2}
       \frac{1}{K} = \frac{1-\delta}{1-\gamma} - \frac{\delta ( 1 + \gamma)}{2 \gamma} \ln \left( 1 - \frac{2 \gamma}{1 + \gamma} \right)  .
    \end{equation}  
    We will approximate this expression in the limit of small $\delta$, which will prove useful when applied to fluid VdW.
    Under this assumption, we simply have $K=1-\gamma$, and $\dPhi$ reads
    \begin{equation} \label{eq.dPhiHT}
        \Delta \Phi_{(HT)} = - \frac{A_\nu (1-\gamma)}{1 + A_\nu} 
    \end{equation}   
    Note the expression of the quantity $\dPhiz$, introduced in equation \eqref{eq.dPhi0}, is simply
    \begin{equation} \label{eq.dPhiHT2}
        \Delta \Phi_0 = - \frac{A_\nu}{1 + A_\nu} ,
    \end{equation}   
    which does not depend on $\gamma$.
    Using $\Delta \Phi_0$ is then found of practical interest for fluid VdW as the parameter $\gamma$ does not need to be introduced and defined for this fluid.

   \subsection{Limiting case}
    In the limit of $\delta \ll  1$, $\Anu \ll 1$ and $\gamma \ll 1$, all fluids exhibit the same expression:
    \begin{equation} \label{eq.dPhi_smallDeltaSmalldPhi}
        \Delta \Phi \sim - \Anu .
    \end{equation}
    This limit is of interest since it corresponds to parameters near the neutral curve when $\delta \ll 1$, since $\Delta \Phi = \mathcal{O}(\delta)$ in this region.  

\bibliographystyle{jfm}
\bibliography{bibliSupercritical}

\end{document}